\documentclass{aa}  

\usepackage{graphicx}
\usepackage{multirow}
\usepackage{txfonts}
\usepackage{hyperref}
\usepackage{lscape}
\bibpunct{(}{)}{;}{a}{}{,} 

\begin{document} 

   \title{Active galactic nuclei cores in infrared-faint radio sources}

   \subtitle{Very long baseline interferometry observations using the Very Long
   Baseline Array}

   \author{A. Herzog\inst{1,2,3}
      \and
     E. Middelberg\inst{1}
	  \and
	  R. P. Norris\inst{3}
	  \and
	  L. R. Spitler\inst{2,4}
	  \and
	  A. T. Deller\inst{5}
	  \and
	  J. D. Collier\inst{6,3}
	  \and
	  Q. A. Parker\inst{2,4}
          }

   \institute{Astronomisches Institut, Ruhr-Universit\"at Bochum, Universit\"atsstr. 150, 44801 Bochum, Germany\\
              \email{herzog@astro.rub.de}
         \and
             Macquarie University, Sydney, NSW 2109, Australia
         \and
            CSIRO Astronomy and Space Science, Marsfield, PO Box 76, Epping, NSW
            1710, Australia
         \and
            Australian Astronomical Observatory, PO Box 915, North Ryde, NSW
            1670, Australia
         \and
            The Netherlands Institute for Radio Astronomy (ASTRON), Postbus 2,
             7990 AA Dwingeloo, The Netherlands
         \and
            University of Western Sydney, Locked Bag 1797, Penrith, NSW 2751,
            Australia}

   \date{Received March 02, 2015; accepted March 31, 2015}

 
  \abstract
   {Infrared-faint radio sources~(IFRS) form a new class of galaxies
   characterised by radio flux densities between tenths and tens of mJy and
   faint or absent infrared counterparts. It has been suggested that these
   objects are radio-loud active galactic nuclei (AGNs) at significant redshifts~($z\gtrsim 2$).}
   {Whereas the high redshifts of IFRS have been recently confirmed based on
   spectroscopic data, the evidence for the presence of AGNs in IFRS
   is mainly indirect. So far, only two AGNs have been unquestionably confirmed
   in IFRS based on very long baseline interferometry~(VLBI) observations. In
   this work, we test the hypothesis that IFRS contain AGNs in a large sample of
   sources using VLBI.}
   {We observed 57~IFRS with the Very Long Baseline Array~(VLBA) down to a
   detection sensitivity in the sub-mJy regime and detected compact cores in
   35~sources.}
   {Our VLBA detections increase the number of VLBI-detected IFRS from 2 to
   37 and provide strong evidence that most---if not all---IFRS contain AGNs. We
   find that IFRS have a marginally higher VLBI detection fraction than
   randomly selected sources with mJy flux densities at arcsec-scales.
   Moreover, our data provide a positive correlation between
   compactness---defined as the ratio of milliarcsec- to arcsec-scale flux
   density---and redshift for IFRS, but suggest a decreasing mean compactness
   with increasing arcsec-scale radio flux density. Based on these findings, we
   suggest that IFRS tend to contain young AGNs whose jets have not formed yet
   or have not expanded, equivalent to very compact objects. We found two IFRS
   that are resolved into two components. The two components are spatially
   separated by a few hundred milliarcseconds in both cases. They might be
   components of one AGN, a binary black hole, or the result of gravitational
   lensing.}
   {}


   \maketitle
%

\section{Introduction}

One of the most controversially discussed topics in current astrophysics is the
question of how active galactic nuclei~(AGNs) influence star forming activity in
their host galaxies. This interaction is known as feedback, and both negative
and positive feedback of the AGN have been proposed.\par

Negative AGN~feedback is thought to be caused by the AGN heating or
disrupting the surrounding gas and, by this, quenching star formation. This scenario consists of two
consecutive phases and may be caused by a merger of two galaxies. In
the first phase, which is called cold mode (or quasar mode; e.g.\
\citealp{Kauffmann2003,Croton2006,Hardcastle2007}), cold gas from the galaxies
involved in the merger fuels the accretion accompanied by high star forming
activity. Finally, the AGN heats or blows away the remaining gas and star
formation ends. This represents the transition to the second phase, the so-called hot mode accretion (also
known as radio mode; e.g.\ \citealp{Narayan1995}). In this phase, the
supermassive black hole~(SMBH) can only accrete hot gas from the halo in
so-called advection dominated accretion flows.\par

The accretion in the cold mode phase is radiatively efficient and
observational evidence is difficult because of the highly obscured nucleus.
In contrast, accretion in the hot mode phase is typically well below the
Eddington limit and the accretion flow radiatively inefficient. In hot
mode, AGN feedback is kinetic and the total energy output usually dominated
by the mechanical power of the radio jets. Radio jets are associated with both
accretion modes and the radio emission can be used to trace the mechanical jet
power (e.g.\ \citealp{Alexander2012} and references
therein).\par

This scenario of negative AGN~feedback is an important ingredient for the
current preferred cosmological $\Lambda$CDM model. In this model, negative
AGN~feedback is needed to make the number of massive and luminous galaxies in
simulations consistent with the observed
number~\citep{Springel2005,Croton2006}.\par

Contrary to this widely accepted negative AGN~feedback, examples have been found
where AGN~activity enhances star formation, i.e.\ showing a positive feedback~(e.g.\
\citealp{Klamer2004,Gaibler2012,Zinn2013,Karouzos2014}). These observations are
explained by the AGN~jets propagating through the gas of the host galaxy, generating
shocks that trigger the gravitational collaps of the gas and, by this, boost
star formation.\par

One important class of object in studying AGN feedback processes is the class of
high-redshift radio galaxies~(HzRGs) since these galaxies can be observed out
to high redshifts. HzRGs are very powerful radio-galaxies that contain
significant emission both from a starburst activity and AGN activity. They are
expected to be the progenitors of the most massive galaxies in the local universe~(e.g.\
\citealp{Seymour2007,deBreuck2010}). Only around 200 of these objects---which
are defined by $z>1$ and $L_\mathrm{3\,GHz} > 10^{26}$\,W\,Hz$^{-1}$---are known
in the entire sky. However, a new class of object has recently been found
that suggests a link to HzRGs: the class of infrared-faint radio sources.\par

\subsection{Discovery and definition of infrared-faint radio sources}
\label{IFRSdefinition}

Infrared-faint radio sources~(IFRS) are characterised by radio emission of the
order of tenths to tens of mJy and associated deep near-infrared faintness.
\citet{Norris2006} and \citet{Middelberg2008ELAIS-S1} discovered these objects
in the deep radio maps of the Australia Telescope Large Area Survey~(ATLAS) in
the Chandra Deep Field South~(CDFS) and the European Large Area IR space
observatory Survey South~1~(ELAIS-S1) as lacking infrared~(IR) counterparts in
the co-located \textit{Spitzer} Wide-area Infrared Extragalactic Survey~(SWIRE;
\citealp{Lonsdale2003}). \citet{Zinn2011} defined two survey-independent
criteria for the selection of IFRS:
\begin{enumerate}[(i)]
  \item $S_{1.4\,\mathrm{GHz}} / S_{3.6\,\mu\mathrm{m}} > 500$ and \item
  $S_{3.6\,\mu\mathrm{m}} < 30\,\mu$Jy.
\end{enumerate}
The high radio-to-IR flux density ratios, ensured by the first criterion, show
that IFRS are clear outliers from the radio-to-IR correlation. The second
criterion is equivalent to a distance selection and prevents ordinary objects at
$z\lesssim 1.4$ from being included in this class of object.\par

Based on the two selection criteria, \citet{Zinn2011} compiled a catalogue of
55~IFRS in the deep fields of the CDFS, ELAIS-S1, \textit{Spitzer} extragalactic
First Look Survey~(xFLS), and the Cosmological Evolution Survey~(COSMOS), based
on the work by \citet{Norris2006}, \citet{Middelberg2008ELAIS-S1}, and
\citet{GarnAlexander2008}. Later, IFRS were also found in the European Large
Area IR space observatory Survey North~1~(ELAIS-N1) field by
\citet{Banfield2011} and in the Lockman Hole field by Maini et al.~(submitted).
Around 100~IFRS have been found in these deep fields, covering a total area of
around 35\,deg$^2$.\par

Recently, \citet{Collier2014}, for the first time, used a different approach and
looked for IFRS in much shallower radio and IR data, which covered a much larger
area compared to the deep fields mentioned above. \citeauthor{Collier2014} used
data from the Unified Radio Catalogue~(URC; \citealp{Kimball2008,Kimball2014}) and from
the all-sky Wide-Field IR Survey Explorer~(WISE; \citealp{Wright2010}). Based on
these data, \citeauthor{Collier2014} compiled a catalogue of 1317~IFRS, all of
them fulfilling both selection criteria from \citet{Zinn2011}.\par

\subsection{The properties of IFRS}

Since the first detection of IFRS by \citet{Norris2006} it has been suggested
that IFRS are radio-loud AGNs at significant redshifts ($z\gtrsim 2$). Different
studies found evidence for this suggestion. \citet{Norris2007} and
\citet{Middelberg2008IFRS_VLBI} presented very long baseline
interferometry~(VLBI) observations of IFRS and detected compact cores in two
IFRS. \citeauthor{Norris2007} used the Australian Long Baseline Array~(LBA) and
targeted two IFRS from the CDFS, out of which one was detected, showing a total
VLBI flux density of 5.0\,mJy at 1.6\,GHz. The source was unresolved on VLBI scales
and its size measured to be less than 0.03\arcsec, corresponding to less than
260\,pc at any redshift. \citeauthor{Middelberg2008IFRS_VLBI} also used
the LBA and observed four IFRS from the ELAIS-S1 field at 1.6\,GHz or 1.4\,GHz.
One out of these four IFRS was detected with a VLBI flux density of
12.5\,mJy at 1.6\,GHz. Based on a flux density of 7\,mJy on the longest
baselines, \citeauthor{Middelberg2008IFRS_VLBI} measured a lower limit
on the brightness temperature of $3.6\times 10^6$\,K. Three IFRS were included
in the sample observed with the Very Long Baseline Array~(VLBA) in the CDFS by
\citet{Middelberg2011VLBA_CDFS}; however, they remained undetected. The two VLBI
detections of IFRS showed that at least a fraction of IFRS contain AGNs.\par

\citet{GarnAlexander2008} and \citet{Huynh2010} found that obscured star forming
galaxies cannot reproduce the characteristics of IFRS because IFRS clearly
deviate from the radio-IR correlation, providing another hint for the AGN
content in IFRS. Furthermore, \citeauthor{GarnAlexander2008} and
\citeauthor{Huynh2010} presented the first spectral energy distribution~(SED)
modelling of IFRS and showed that 3C sources like 3C\,273 are in agreement with
the characteristics of IFRS if these sources are at redshifts~$z\gtrsim 2$.
\citet{Herzog2014} showed that the SED of their sample of IFRS can only be
explained by radio-loud AGN templates.\par

A potential link between IFRS and HzRGs was first suggested by \citet{Huynh2010}
based on the similarly high radio-to-IR flux density ratios.
\citet{Middelberg2011} showed that IFRS have steeper radio spectra\footnote{The
spectral index is defined as $S\propto \nu^\alpha$.} (median spectral index
$\alpha = -1.4$) than the general radio source population~($\alpha = -0.86$) and
the AGN source population~($\alpha = -0.82$), using data in the ELAIS-S1 field
between 2.3\,GHz and 8.4\,GHz. Moreover, \citeauthor{Middelberg2011} found that
the radio spectrum of IFRS is even steeper than that of HzRGs ($\alpha =
-1.02$). \citet{Norris2011} pointed out that HzRGs are the only objects known at
high redshifts that share the extreme radio-to-IR flux density ratios with IFRS.
Based on this similarity and the deep IR faintness of IFRS,
\citeauthor{Norris2011} suggested that IFRS might follow the correlation between
$3.6\,\mu$m flux density and redshift found for HzRGs~\citep{Seymour2007},
similar to the $K-z$ relation~\citep{Willott2003}. Herzog at al.\ (submitted) showed that the
non-detection of IFRS in deep far-IR \textit{Herschel} observations can only be
explained by SED templates of HzRGs.\par

\citet{Collier2014} and \citet{Herzog2014} presented the first spectroscopic
redshifts of IFRS and found all of 22 but one---which is most likely a
misidentification or a star forming galaxy with an AGN---redshifts in the range
$1.7\leq z \leq 3.0$, confirming the suggested high-redshift nature of IFRS.
Furthermore, both studies found their data in agreement with the suggested
correlation between near-IR flux density and redshift, indicating that most IFRS
in deep fields---which were summarised by \citet{Zinn2011} and Maini et
al.~(submitted)---might be at even higher redshifts.

\subsection{Populations of IFRS found in deep and shallow surveys}

Here, we consider the relationship between IFRS found in deep surveys of small
area and those found in shallow all-sky surveys. First, all IFRS found in the
various works mentioned above fulfil the---in some cases slightly
changed---selection criteria by \citet{Zinn2011} and therefore qualify as IFRS.
Maini et al.~(submitted) lowered the radio-to-IR flux density criterion and
replaced the IR flux density criterion by an extension criterion.
Others---like \citet{Norris2007} and \citet{Middelberg2008ELAIS-S1}---required
IFRS to be undetected at 3.6\,$\mu$m without applying a radio-to-IR flux
density criterion. Nevertheless, these IFRS, which were selected based
on slightly different criteria, are close to fulfil the criteria by
\citeauthor{Zinn2011}.\par

While the IFRS in deep fields usually have 1.4\,GHz flux densities of
tenths of mJy to a few mJy, the IFRS in the all-sky survey have median radio
flux densities of several tens of mJy, some reaching even several hundred mJy.
Similarly, each IFRS in the sample from \citet{Collier2014} provides an IR
counterpart at 3.4\,$\mu$m with a mean flux density of around 25\,$\mu$Jy, while
a significant fraction of IFRS in deep fields has no IR counterpart in the even
deeper 3.6\,$\mu$m data. Thus, the median IFRS in deep fields is
both radio and IR-fainter than the median IFRS in the shallower all-sky sample
from \citeauthor{Collier2014}.\par

All spectroscopic redshifts of IFRS from both the deep fields and the shallow
survey were found to be in the same redshift range $1.7\lesssim z \lesssim 3.0$.
However, a selection effect putatively biases the observed spectroscopic
redshift distribution of IFRS in the deep fields since the IFRS with known
spectroscopic redshifts in these fields are the optically and IR brightest IFRS
in that sample. Since their optical, IR, and radio properties are similar to
those of the all-sky IFRS, it is expected that these IFRS represent the overlap
between the fainter IFRS population in deep fields and the brighter all-sky IFRS
population~\citep{Herzog2014}.\par

It has been suggested by \citet{Collier2014} that their IFRS sample consists of
the lowest-redshift IFRS while the IFRS found in deep fields are on average at
higher redshifts. This suggestion is in agreement with the correlation between
3.6\,$\mu$m flux density and redshift discussed by \citet{Norris2011},
\citeauthor{Collier2014}, and \citet{Herzog2014}. The overlapping spectroscopic
redshifts found for both subsets are in agreement with this suggested
unification, too.\par

\citet{Collier2014} could only set a lower limit of $\sim
0.1\,\mathrm{deg}^{-2}$ for the sky density of IFRS with $S_\mathrm{1.4\,GHz}
\geq 7.5$\,mJy because of the non-uniform sensitivity of the WISE survey. 
In contrast, the sky density of IFRS in deep fields is of the order of a few per
square degree and might reach 30\,deg$^{-2}$~\citep{Zinn2011}.\par

In this paper, we test the AGN content in IFRS based on VLBI observations with
the VLBA of a large number of sources taken from the all-sky catalogue of
IFRS~\citep{Collier2014}. In Sect.~\ref{sample_observations}, we describe our
sample and the observing strategy. We discuss data calibration, imaging, flux
measuring, and redshifts in Sect.~\ref{calibration_opticalproperties_redshifts}.
We analyse our data in Sect.~\ref{analysis} with respect to detection fraction
(Sect.~\ref{detection_fraction}), compactness (Sect.~\ref{compactness}), and
individual sources (Sect.~\ref{individual_sources}). We discuss the implications
of our analysis in Sect.~\ref{discussion}, and present our conclusions in
Sect.~\ref{conclusions}. Throughout this paper, we use flat $\Lambda$CDM
cosmological parameters $\Omega_\Lambda = 0.7$, $\Omega_\textrm{M} = 0.3$, $H_0
= 70$~km~s$^{-1}$~Mpc$^{-1}$ and the calculator by \citet{Wright2006}. We quote
$1\sigma$ confidence intervals of binomial population proportions based on the
bayesian approach, following \citet{Cameron2011_beta}.\par

\section{Sample and observations}
\label{sample_observations}
We selected all IFRS from the catalogue from \citet{Collier2014} which were
located within 1\,deg of a VLBA calibrator. This low angular separation between source
and calibrator ensured the phase coherence required for VLBI observations. Since
1.4\,GHz VLBI observations of the calibrators were not available, we required
the calibrators to have a 2.3\,GHz flux density of at least 0.2\,Jy on a baseline of
5000\,km. Out of the 1317~IFRS presented by \citeauthor{Collier2014}, 110 were
found to provide a calibrator which fulfills the given conditions.\par

A VLBI detection provides unambiguous evidence for an AGN because compact radio
emission in AGNs is a non-thermal process and results in brightness temperatures
of more than $10^6$\,K to which our observations are sensitive. In contrast,
compact radio emission in starburst galaxies, which is usually dominated by
thermal free-free emission, is represented by brightness temperatures of around
$10^4$\,K~(e.g.\ \citealp{Condon1991}). Although brightness temperatures of
$10^6$\,K can also be produced by very luminous radio supernovae~(SNe;
\citealp{Huang1994,Smith1998}), \citet{Kewley2000} showed that the probability
for a VLBI-detected radio SN in a galaxy sample is very low. Therefore, a VLBI
detection provides strong evidence for an AGN. However, it should be noted that the reverse is not
true, i.e.\ the non-detection of a source in a VLBI observation does not imply
the non-existence of an AGN. Instead, a VLBI non-detection implies significant
extended emission compared to the compact core. The ratio of extended
emission to core emission depends on beaming which can boost or suppress
the compact core emission, AGN age, and the surrounding medium, affecting
the brightness and extent of the diffuse radio lobes.\par

Since this is a detection experiment, the $uv$-coverage is not critically
important and a minimum number of six out of ten VLBA antennas was requested.
Since the individual observations were short and independent of the weather
conditions because of the observing frequency of 1.4\,GHz, the observations were
scheduled in filler time. Although the maps resulting from the data will
be of rather poor quality, they will unambiguously resolve potential compact
components.\par

Out of 110~proposed objects, 57~IFRS were observed in semester~14A in
project~BH197. The 57~observed IFRS were randomly selected based on the
IFRS positions and available filler time at the VLBA. These observed
sources have 1.4\,GHz integrated flux densities between 11\,mJy and 183\,mJy in
the NRAO VLA Sky Survey~(NVSS; \citealp{Condon1998}). The VLBA observations were
set to a bandwidth of 32\,MHz in each of the eight basebands which were observed
in dual polarisation at 1.4\,GHz, resulting in a total data rate of
2048\,Mbps.\par

Each of the 57~epochs had a total observing time of one hour. We decided to use
two different approaches for the scan settings, depending on the distance
between source and calibrator. If the separation between source and calibrator
was less than 25\arcmin, we continuously pointed at the position in between IFRS
and calibrator to prevent unnecessary nodding between the two sources. If the
separation was more than 25\arcmin, we alternately observed the calibrator for
60\,s and the source for 225\,s, starting and ending with a scan on the
calibrator. The resulting observing time on the IFRS was around 45\,min. The
data were correlated using the VLBA Distributed FX (DiFX) software correlator
\citep{Deller2007,Deller2011}.

\section{VLBA data calibration, optical properties, and redshifts}
\label{calibration_opticalproperties_redshifts}

\subsection{Data calibration, imaging, and flux measurement of the VLBA data}
\label{calibration_imaging}

Calibration and imaging of the individual epochs was carried out based on a
ParselTongue script. ParselTongue~\citep{Kettenis2006} is a Python-based
interface to the Astronomical Image and Processing
System\footnote{\url{http://www.aips.nrao.edu/}}~(AIPS). The
calibration and imaging strategy used here is very similar to the procedure
described by \citet{Deller2014}. However, we calibrated the amplitudes using the
technique suggested by VLBA Scientific Memo \#37 (R.~Craig Walker; Dec 15, 2014).\par

In the pipeline, we loaded the data into AIPS and applied a priori flags and
manual flags. We then corrected for ionospheric effects using the task
\texttt{TECOR}, applied the latest earth orientation parameters, and corrected
for parallactic angles, in the latter cases using the task \texttt{CLCOR}.
Amplitudes were calibrated based on autocorrelation data using the task
\texttt{ACCOR}. We corrected for primary beam effects using the task
\texttt{CLVLB}, following the procedure outlined by \citet{Middelberg2013}.
Delay correction was carried out based on the calibrator, using the task
\texttt{FRING} and a solution interval of 2\,min. We applied a bandpass
correction using the task \texttt{BPASS}. The new task \texttt{ACSCL},
implemented in AIPS in consequence of the flux density calibration errors
described in the VLBA Scientific Memo \#37, kept the calibrated autocorrelation
values at unity. The flux density calibration was completed by using
\texttt{APCAL}, calibrating the amplitudes based on system temperatures and
gains. We used the AIPS/ParselTongue implementation of
Pieflag~\citep{Middelberg2006}, \textit{dynspec-flagger}, to automatically flag
data affected by interference. Using the task \texttt{CALIB}, we performed one
iteration of phase and amplitude self-calibration on the calibrator and applied
the solution on the target. Finally, we imaged the UV~data based on the task
\texttt{IMAGR} into a map of 2048$\times$2048~pixels with a pixel size of
1\,milliarcsecond~(mas)---matching the angular resolution of $\sim
5$\,mas---using uniform weighting. In this step, we cleaned the dirty image in a
given box, down to a flux of three times the rms or up to 500~iterations.\par

\begin{figure}
	\centering
		\includegraphics[width=\hsize]{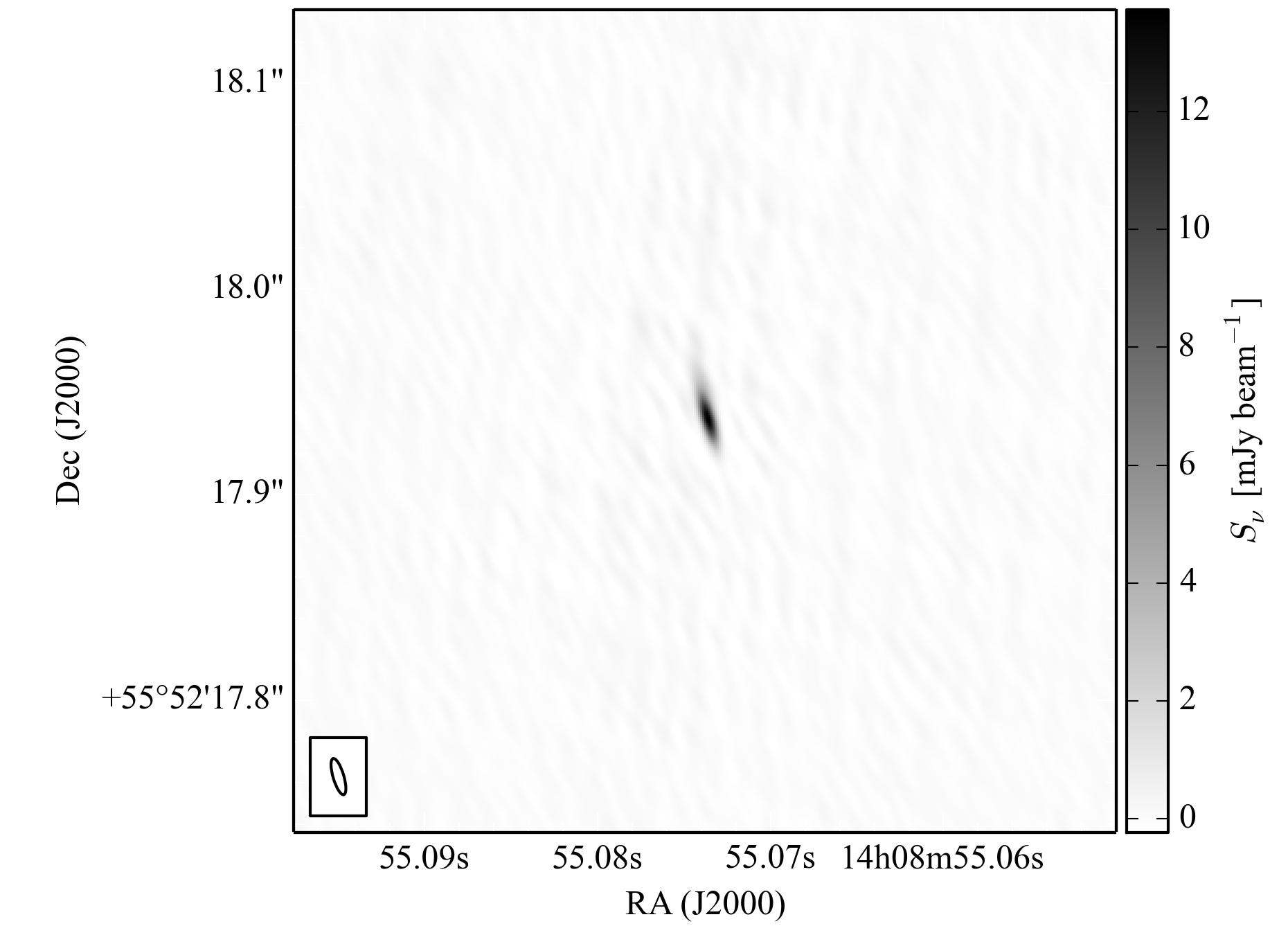}
		\caption{VLBA map of IFRS~F0398. The source is slightly resolved with a
		peak flux density of 13.7\,mJy\,$\mathrm{beam}^{-1}$ and an integrated flux
		density of 17.2\,mJy.}
	\label{fig:F0398}
\end{figure}

The mean synthesised beam size in our observations is $14.3\times
4.7\,\mathrm{mas}^2$. Because the linear scale is limited in a
$\Lambda$CDM cosmology at redshifts $0.5\leq z \leq 12$ between 4\,kpc/\arcsec
and 8.5\,kpc/\arcsec, we were able to convert this beam size over this redshift
range to an area of between $57\,\mathrm{pc} \times 19\,\mathrm{pc}$ and
$122\,\mathrm{pc} \times 40\,\mathrm{pc}$. The rms in the final maps is around
60\,$\mu$Jy\,beam$^{-1}$, depending on the number of antennas. Of the
57~observations, 31, 21, 4, and 1 were carried out with 10, 9, 8, and
7~antennas, respectively. Since the longest VLBA baseline of 8611\,km---between
the stations Mauna Kea and St.~Croix---was available in all observations, the
east-west angular resolution of around 5\,mas is similar for all
57~observations, whereas the north-south angular resolution varies
slightly because of changing antenna availability.
As an example, Fig.~\ref{fig:F0398} shows the final map of IFRS~F0398.
\par

We measured the flux density of each component using the AIPS task
\texttt{JMFIT}, setting a box of size of 40~pixels around the components. For a
few sources, image artefacts resulted in unreasonable fluxes. In these cases, we
manually measured flux densities using the task \texttt{TVSTAT}.
Checks on control sources resulted in consistent flux measurements based on
\texttt{TVSTAT} and \texttt{JMFIT}. For the brighter component in IFRS F0030,
we measured the flux density using the \texttt{blobcat}
package~\citep{Hales2012} because of the complex structure of this source. Following \citet{Deller2014},
we set a $6.75\sigma$ detection limit for all sources, corresponding to a mean
detection sensitivity of around 450\,$\mu$Jy\,beam$^{-1}$. Most of our sources
are slightly resolved, but insufficiently resolved to determine the morphology.
The resulting flux densities und flux density upper limits are summarised in
Table~\ref{tab:catalogue}.
\begin{longtab}
\begin{landscape}
\centering
\begin{longtable}{c c c c c c c c c c}
    \caption{\label{tab:catalogue} Component catalogue of 57~IFRS observed with
    the VLBA. Listed is the identifier, the NVSS ID, the FIRST
    position, the NVSS peak and integrated flux densities, the VLBA peak flux
    density and integrated flux density, the VLBA S/N, and the VLBA beam size. If a source is found to be composed of several components in the VLBA observations, we list the components individually. In case of a
    non-detection, we quote a $6.75\sigma$~peak flux density upper limit. The
    identifier follows the numbering by \citet{Collier2014}. NVSS and
    FIRST data are taken from \citeauthor{Collier2014}.}\\
    \hline \hline
    Identifier & NVSS ID & RA$_\mathrm{FIRST}$ & Dec$_\mathrm{FIRST}$ &
    $S_{\mathrm{NVSS,\,peak}}$ & $S_{\mathrm{NVSS,\,int}}$ & $S_{\mathrm{VLBA,\,peak}}$ & $S_{\mathrm{VLBA,\,int}}$ &
    S/N & $\Theta_\mathrm{maj} \times \Theta_\mathrm{min}$ \\ 
    & & J2000.0 & J2000.0 & [mJy\,beam$^{-1}$] & [mJy] &
    [mJy\,beam$^{-1}$] & [mJy] & & [mas\,$\times$\,mas] \\
    \hline
    \endfirsthead
    \multicolumn{10}{c}{\tablename\ \thetable\ -- \textit{Continued}} \\
    \hline \hline
    Identifier & NVSS ID & RA$_\mathrm{FIRST}$ & Dec$_\mathrm{FIRST}$ &
    $S_{\mathrm{NVSS,\,peak}}$ & $S_{\mathrm{NVSS,\,int}}$ & $S_{\mathrm{VLBA,\,peak}}$ & $S_{\mathrm{VLBA,\,int}}$ &
    S/N & $\Theta_\mathrm{maj} \times \Theta_\mathrm{min}$ \\
    & & J2000.0 & J2000.0 & [mJy\,beam$^{-1}$] & [mJy] &
    [mJy\,beam$^{-1}$] & [mJy] & & [mas\,$\times$\,mas] \\
    \hline
    \endhead
    \hline
    \endfoot
    \hline
F0013 & NVSS J014418--092158 & 01:44:18.196 & --09:21:54.77 & 13.3 & 13.3 & <0.3
& -- & -- & -- \\
F0030\_1 & NVSS J021557--082517 & 02:15:57.080 & --08:25:17.55 & 70.7 & 75.6 &
4.1 & 21.8 & 54.9 & 14.0\,$\times$\,4.4 \\
F0030\_2 & NVSS J021557--082517 & 02:15:57.080 & --08:25:17.55 & 70.7 & 75.6 &
3.5 & 6.0 & 45.7 & 14.0\,$\times$\,4.4 \\
F0037 & NVSS J022022--011017 & 02:20:22.049 & --01:10:16.49 & 30.0 & 30.6 & 1.3
& 4.8 & 27.9 & 13.4\,$\times$\,4.6 \\
F0052 & NVSS J023033--030909 & 02:30:33.435 & --03:09:08.37 & 49.7 & 50.5 & <0.6
& -- & -- & -- \\
F0072 & NVSS J024150--032011 & 02:41:50.196 & --03:20:12.17 & 19.2 & 19.2 & 6.5
& 11.1 & 64.4 & 13.7\,$\times$\,4.7 \\
F0076 & NVSS J024346--050737 & 02:43:46.908 & --05:07:36.51 & 42.1 & 43.3 & 5.5
& 21.2 & 74.8 & 20.7\,$\times$\,4.7 \\
F0081 & NVSS J024700+062831 & 02:47:00.457 & +06:28:34.34 & 28.7 & 29.3 & <0.3 & -- & -- & -- \\ 
F0100 & NVSS J090337+515142 & 09:03:37.302 & +51:51:42.90 & 19.7 & 19.7 & 10.3 & 11.2 & 263.8 & 14.0\,$\times$\,4.2 \\ 
F0106 & NVSS J093243+521400 & 09:32:43.626 & +52:13:59.53 & 13.0 & 13.3 & 0.7 & 1.1 & 17.4 & 13.7\,$\times$\,4.8 \\ 
F0127 & NVSS J104946+531947 & 10:49:46.291 & +53:19:50.81 & 10.9 & 15.5 & <0.4 & -- & -- & -- \\ 
F0146 & NVSS J112624+375323 & 11:26:23.726 & +37:53:34.75 & 8.4 & 16.1 & 3.6 & 3.9 & 74.8 & 12.4\,$\times$\,6.0 \\ 
F0149 & NVSS J113330+585506 & 11:33:29.982 & +58:55:05.01 & 39.4 & 39.6 & 24.0 & 29.3 & 93.3 & 16.8\,$\times$\,4.6 \\ 
F0154 & NVSS J114333+582209 & 11:43:33.395 & +58:22:07.59 & 15.7 & 16.5 & 3.7 & 7.1 & 83.6 & 13.2\,$\times$\,4.5 \\ 
F0167 & NVSS J115544+495437 & 11:55:44.856 & +49:54:36.55 & 53.8 & 54.2 & <0.4 & -- & -- & -- \\ 
F0169 & NVSS J115900+474005 & 11:59:00.500 & +47:40:05.26 & 17.5 & 17.7 & <0.2 & -- & -- & -- \\ 
F0173 & NVSS J120221+482514 & 12:02:21.184 & +48:25:13.80 & 58.4 & 59.9 & 26.4 & 40.4 & 248.7 & 19.8\,$\times$\,4.7 \\ 
F0187 & NVSS J121731+485954 & 12:17:31.830 & +48:59:53.73 & 43.9 & 44.1 & <0.4 & -- & -- & -- \\ 
F0189 & NVSS J121814+590516 & 12:18:14.097 & +59:05:16.16 & 26.9 & 27.1 & 0.7 & 1.0 & 15.2 & 15.8\,$\times$\,4.1 \\ 
F0194 & NVSS J122524+433439 & 12:25:24.786 & +43:34:38.92 & 25.7 & 25.7 & <0.3 & -- & -- & -- \\ 
F0197 & NVSS J122743+364252 & 12:27:43.502 & +36:42:55.83 & 23.8 & 24.1 & <0.3 & -- & -- & -- \\ 
F0209 & NVSS J123952+604958 & 12:39:52.632 & +60:49:55.77 & 18.9 & 19.5 & 3.0 & 7.1 & 71.2 & 14.3\,$\times$\,6.1 \\ 
F0222 & NVSS J125148--064218 & 12:51:48.680 & --06:42:17.01 & 11.6 & 23.9 & <0.3
& -- & -- & -- \\
F0241 & NVSS J130316+481558 & 13:03:16.442 & +48:15:57.70 & 13.8 & 13.9 & 6.9 & 8.1 & 137.8 & 12.8\,$\times$\,5.7 \\ 
F0244 & NVSS J130748+555452 & 13:07:48.324 & +55:54:50.53 & 14.2 & 14.5 & 1.0 & 1.3 & 22.4 & 13.8\,$\times$\,5.0 \\ 
F0251 & NVSS J131322+322105 & 13:13:22.589 & +32:21:10.13 & 36.1 & 37.6 & <0.4 & -- & -- & -- \\ 
F0257\_1 & NVSS J131551+512710 & 13:15:51.150 & +51:27:10.01 & 36.0 & 36.7 &
2.5 & 5.7 & 77.3 & 14.2\,$\times$\,4.4 \\
F0257\_2 & NVSS J131551+512710 & 13:15:51.150 & +51:27:10.01 & 36.0 & 36.7 &
1.2 & 2.1 & 35.9 & 14.2\,$\times$\,4.4 \\
F0273 & NVSS J132804+431418 & 13:28:04.382 & +43:14:17.27 & 18.1 & 18.5 & 2.2 & 2.3 & 39.3 & 12.7\,$\times$\,4.2 \\ 
F0277 & NVSS J133024+221800 & 13:30:24.695 & +22:18:00.36 & 177.0 & 179.6 & 7.8 & 20.3 & 104.4 & 12.3\,$\times$\,4.8 \\ 
F0283 & NVSS J133431+543930 & 13:34:31.807 & +54:39:32.23 & 13.5 & 14.4 & <0.3 & -- & -- & -- \\ 
F0293 & NVSS J133733+591837 & 13:37:33.098 & +59:18:37.68 & 44.7 & 45.4 & 13.6 & 21.2 & 214.3 & 18.3\,$\times$\,4.1 \\ 
F0319 & NVSS J134921+081217 & 13:49:21.396 & +08:12:15.75 & 27.2 & 27.9 & <0.4 & -- & -- & -- \\ 
F0334 & NVSS J135248+093020 & 13:52:48.328 & +09:30:16.60 & 16.9 & 17.4 & <0.4 & -- & -- & -- \\ 
F0351 & NVSS J135601--012539 & 13:56:01.445 & --01:25:38.70 & 34.9 & 35.3 & 3.0
& 4.2 & 51.0 & 13.0\,$\times$\,4.7 \\
F0382 & NVSS J140707+285558 & 14:07:07.241 & +28:55:56.39 & 66.7 & 68.8 & <0.3 & -- & -- & -- \\ 
F0385 & NVSS J140730+040234 & 14:07:30.593 & +04:02:34.61 & 16.3 & 16.5 & 3.1 & 4.8 & 55.8 & 18.1\,$\times$\,4.2 \\ 
F0398 & NVSS J140855+555218 & 14:08:54.995 & +55:52:17.62 & 62.4 & 62.9 & 13.7 & 17.2 & 184.1 & 18.1\,$\times$\,5.2 \\ 
F0406 & NVSS J141004+024051 & 14:10:04.764 & +02:40:49.82 & 30.0 & 32.9 & 19.3 & 21.9 & 219.2 & 13.6\,$\times$\,4.7 \\ 
F0471 & NVSS J142228+264716 & 14:22:28.909 & +26:47:16.63 & 19.8 & 19.8 & 5.1 & 5.0 & 136.9 & 12.5\,$\times$\,4.8 \\ 
F0472 & NVSS J142241+363956 & 14:22:41.669 & +36:39:57.74 & 17.1 & 18.2 & 0.7 & 0.6 & 14.2 & 13.4\,$\times$\,4.5 \\ 
F0509 & NVSS J143110+360317 & 14:31:10.868 & +36:03:17.05 & 16.3 & 18.3 & <0.2 & -- & -- & -- \\ 
F0588 & NVSS J144500+624605 & 14:45:00.769 & +62:46:05.55 & 16.7 & 17.2 & <0.4 & -- & -- & -- \\ 
F0611 & NVSS J144924+085628 & 14:49:24.799 & +08:56:32.59 & 36.1 & 36.8 & <0.4 & -- & -- & -- \\ 
F0633 & NVSS J145334--014513 & 14:53:34.184 & --01:45:13.48 & 18.0 & 18.1 & 0.4
& 2.6 & 9.5 & 13.0\,$\times$\,4.9 \\
F0726 & NVSS J150623+103048 & 15:06:23.301 & +10:30:47.54 & 49.0 & 50.0 & 20.8 & 32.0 & 317.8 & 12.4\,$\times$\,4.3 \\ 
F0732 & NVSS J150649+422059 & 15:06:49.247 & +42:20:59.02 & 101.8 & 103.6 & 9.1 & 25.5 & 167.3 & 13.8\,$\times$\,5.1 \\ 
F0787 & NVSS J151557+201248 & 15:15:57.840 & +20:12:47.13 & 17.5 & 18.0 & 2.6 & 3.2 & 56.6 & 12.8\,$\times$\,4.3 \\ 
F0807 & NVSS J151817+042327 & 15:18:17.648 & +04:23:26.84 & 77.5 & 78.8 & 6.5 & 13.5 & 58.8 & 15.8\,$\times$\,4.2 \\ 
F0838 & NVSS J152348+321541 & 15:23:48.347 & +32:15:43.79 & 96.5 & 98.8 & <0.3 & -- & -- & -- \\ 
F0912 & NVSS J153826+145505 & 15:38:26.812 & +14:55:05.91 & 14.7 & 15.1 & <0.6 & -- & -- & -- \\ 
F1037 & NVSS J160235+310832 & 16:02:35.685 & +31:08:33.12 & 13.9 & 14.2 & 1.3 & 1.4 & 25.6 & 16.0\,$\times$\,4.1 \\ 
F1111 & NVSS J161910+483709 & 16:19:10.025 & +48:37:13.21 & 12.2 & 12.6 & 8.5 & 10.9 & 199.4 & 14.0\,$\times$\,4.3 \\ 
F1268 & NVSS J172102+333445 & 17:21:02.788 & +33:34:47.29 & 25.7 & 26.7 & <0.3 & -- & -- & -- \\ 
F1286 & NVSS J172923+390532 & 17:29:24.122 & +39:05:31.52 & 11.6 & 11.6 & <0.3 & -- & -- & -- \\ 
F1287 & NVSS J173019+460128 & 17:30:19.022 & +46:01:28.48 & 43.4 & 43.9 & 18.4 & 28.5 & 309.9 & 14.5\,$\times$\,3.9 \\ 
F1301 & NVSS J173517+474300 & 17:35:17.371 & +47:42:59.30 & 180.0 & 183.8 & 0.9 & 2.6 & 14.2 & 12.0\,$\times$\,6.6 \\ 
F1305 & NVSS J173703+494446 & 17:37:02.943 & +49:44:46.26 & 23.0 & 23.4 & 5.7 & 7.9 & 71.3 & 14.9\,$\times$\,3.9 \\ 
F1313 & NVSS J174243+621908 & 17:42:43.642 & +62:19:08.65 & 79.6 & 82.5 & 9.2 & 29.9 & 176.2 & 15.2\,$\times$\,3.7 \\ 
    \hline
    \end{longtable}
    \end{landscape}
\end{longtab}
Since observations and data calibration are very similar to the approach taken
by \citeauthor{Deller2014} who targeted more than 20\,000 sources in their mJy
Imaging VLBA Exploration~(mJIVE) survey, we expect our flux densities to be of
similar accuracy of 20\%.

\subsection{Optical properties and redshifts}
\label{opticalproperties_redshifts}
Our sample was taken from the all-sky IFRS catalogue from \citet{Collier2014}
who cross-matched their sources with the Sloan Digital Sky Survey~(SDSS)
DR9~\citep{Ahn2012}. For the purpose of this work, we cross-matched our
sample of VLBA-observed IFRS to the recent SDSS DR10~\citep{Ahn2014}.
53~($93^{+2}_{-5}$\%) out of our IFRS are covered by SDSS~DR10.
Eleven~($21^{+7}_{-4}$\%) out of these 53~IFRS provide photometric counterparts
which are all close to the sensitivity limit of SDSS.\par

\begin{figure*}
	\centering
		\includegraphics[width=8.5cm]{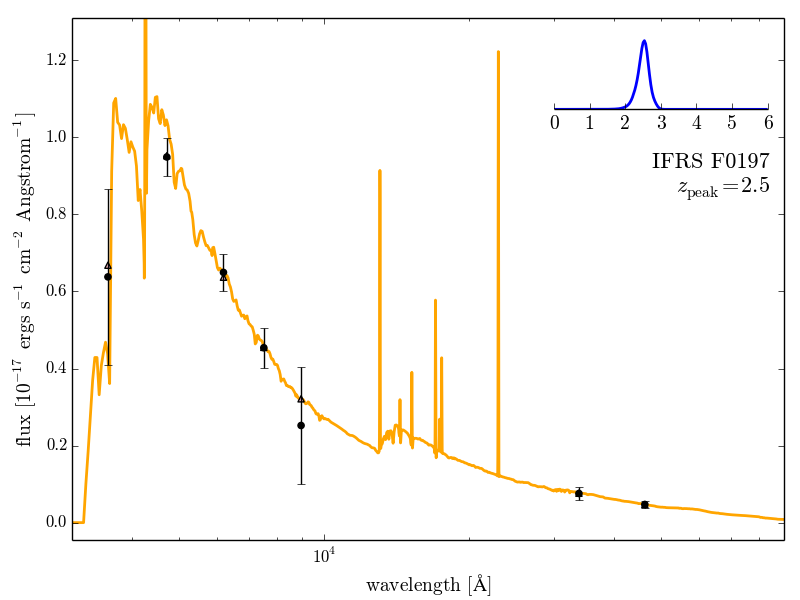}
		\includegraphics[width=8.5cm]{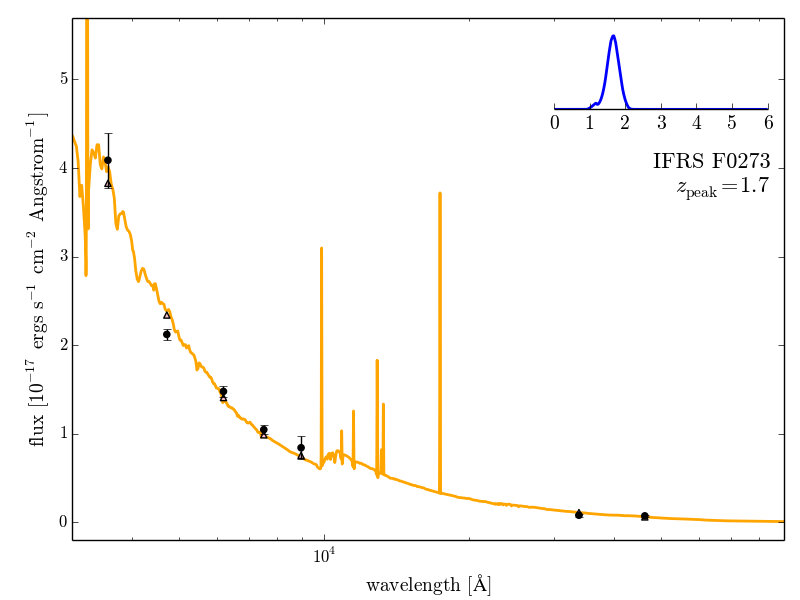}
		\caption{Resulting photometric redshift fit from \textit{EAZY} for IFRS F0197
		(left) and F0273 (right). The SED (orange line)
		shows the best fit template based on the peak redshift in the
		observer's frame.
		Black dots with error bars represent photometric data points of the IFRS,
		whereas black triangles show the flux density of the fitted SED at the same wavelengths. The
		subplots in the upper right of both plots show the redshift-probability
		distribution. The redshift of the peak in the probability distribution is
		quoted below each subplot.}
	\label{fig:photzfit}
\end{figure*}

\begin{figure}
	\centering
		\includegraphics[width=\hsize]{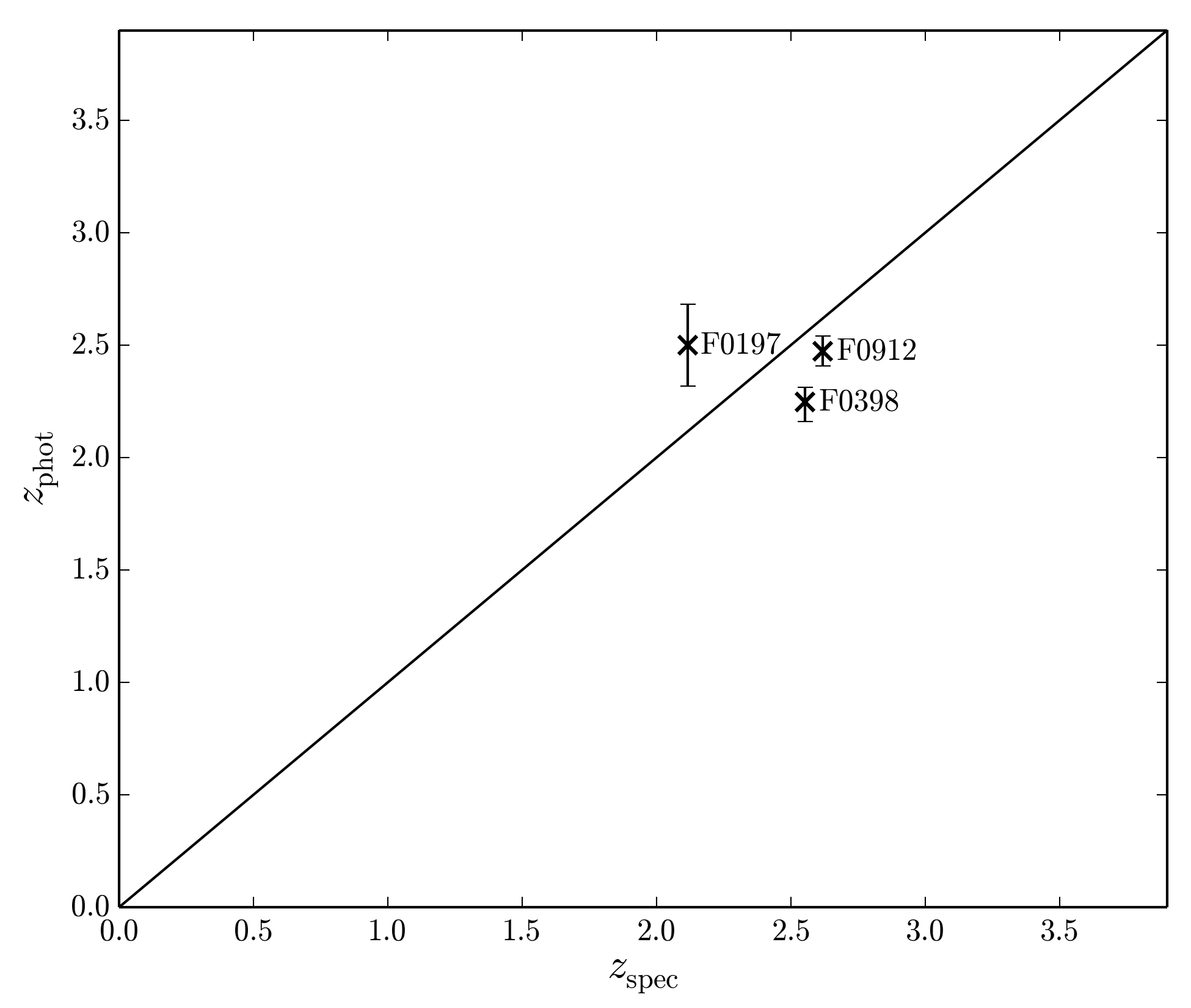}
		\caption{Comparison of the photometric redshifts obtained from \textit{EAZY}
		and the spectroscopic redshifts from SDSS DR10. The error bars
		show the $1\sigma$ uncertainties in the photometric redshifts as determined by
		\textit{EAZY}. The solid line represents the positions of exact agreement
		between photometric and spectroscopic redshifts.}
	\label{fig:redshifts}
\end{figure}

We used the software \textit{EAZY}~\citep{Brammer2008} with the standard
template set to derive photometric redshifts for our sources. Since most IFRS in
the catalogue from \citet{Collier2014} are only detected in the WISE bands W1
(3.4\,$\mu$m) and W2 (4.6\,$\mu$m)---apart from the radio detections which are
not used in the redshift fitting---measuring photometric redshifts is impossible for most of the IFRS in
our sample. However, measuring photometric redshifts is possible for those IFRS
with SDSS counterparts. For these eleven IFRS in our sample with SDSS counterparts,
we obtained ten photometric redshifts using \textit{EAZY}. As examples, we show
the resulting fits for IFRS F0197 and F0273 in Fig.~\ref{fig:photzfit}.
The fitting of IFRS~F0277 failed. Table~\ref{tab:redshifts} summarises the
photometric redshifts and---where applicable---spectroscopic redshifts for our
subsample of IFRS with SDSS DR10 counterparts.
\begin{table}
\renewcommand{\arraystretch}{1.3}
	\caption{Redshift information for those IFRS with SDSS DR10 detections. Listed
	is the IFRS ID, the $u$ band model magnitude from SDSS DR10, the photometric
	redshift measured in this work using \textit{EAZY}, and the spectroscopic
	redshift from SDSS DR10. The SED fitting for F0277 failed.}
	\label{tab:redshifts}
 \centering
 \begin{tabular}{c|c|c|c}
\hline \hline
  IFRS & $u$ & $z_\mathrm{phot}$ & $z_\mathrm{spec}$ \\
  ID & [mag] & & \\
\hline
  F0146 & $20.21 \pm 0.05$ & $1.26^{+0.29}_{-0.25}$ & -- \\
  F0194 & $23.82 \pm 0.95$ & $0.78^{+0.42}_{-0.34}$ & -- \\
  F0197 & $22.83 \pm 0.32$ & $2.54^{+0.14}_{-0.22}$ & $2.1150 \pm 0.0014$ \\
  F0273 & $20.84 \pm 0.08$ & $1.65^{+0.20}_{-0.18}$ & -- \\
  F0277 & $24.16 \pm 0.86$ & -- & -- \\
  F0293 & $22.94 \pm 0.33$ & $3.02^{+0.09}_{-0.08}$ & -- \\
  F0398 & $20.49 \pm 0.06$ & $2.24^{+0.07}_{-0.08}$ & $2.55265 \pm 0.00021$ \\
  F0726 & $23.21 \pm 0.52$ & $3.07^{+0.26}_{-0.24}$ & -- \\
  F0732 & $22.88 \pm 0.27$ & $2.26^{+0.27}_{-0.28}$ & -- \\
  F0912 & $21.55 \pm 0.10$ & $2.47^{+0.06}_{-0.06}$ & $2.61873 \pm 0.00023$ \\
  F1037 & $25.14 \pm 0.79$ & $0.39^{+3.77}_{-0.08}$ & -- \\
\hline
\end{tabular}
\end{table}
Fig.~\ref{fig:redshifts} shows the photometric redshifts as a function of the
spectroscopic redshifts for the three IFRS in our sample for which SDSS DR10 provides
spectroscopic redshifts.
\par

Based on the SEDs resulting from the fitting of photometric redshifts, we find
that most of our fitted IFRS are very blue. These characteristics can be
explained by a type~I AGN in the optical. However, we stress that we might be
significantly affected by selection bias since these IFRS are very close to the
detection sensitivity of SDSS and SDSS is more sensitive to blue objects.
Nevertheless, this study shows that at least some IFRS are very blue, non-dusty
galaxies.

\section{Analysis}
\label{analysis}

\subsection{VLBI detection fraction}
\label{detection_fraction}
In our VLBA observations, we detected 35~($61^{+6}_{-7}$\%) out of
57~observed IFRS, showing peak flux densities between 0.4\,mJy\,beam$^{-1}$ and 26.4\,mJy\,beam$^{-1}$
and integrated flux densities between 0.6\,mJy and 40.4\,mJy as listed in
Table~\ref{tab:catalogue}.\par

The detection of a source in VLBI observations with brightness temperatures
above $10^6$\,K---which is reached for our VLBA observations---is an unambiguous
sign for an AGN as discussed in Sect.~\ref{sample_observations}. However, we
note that the reverse is not true, i.e.\ the non-detection of a source in VLBI
observations does not exclude the existence of an AGN. The detection of 35 out
of 57~IFRS in our VLBA observations provides strong evidence that most---if not
all---IFRS contain AGNs.\par

\subsubsection{VLBI detection fraction compared to other samples}
\label{detection_fraction_AGN}
The detection fraction in our VLBA observations of $61^{+6}_{-7}$\% down to a
$6.75\sigma$ detection limit of $\sim 0.45$\,mJy\,beam$^{-1}$ is significantly
higher than the detection fractions found by \citet{Garrington1999} or
\citet{Deller2014} who targeted large samples of sources from the Faint Images
of the Radio Sky at Twenty centimetres (FIRST; \citealp{Becker1995}) survey in
VLBI observations and detected 35\% and 20\%, respectively. However, the
sensitivity of the respective VLBI observations and the sample selection
criteria were different. \citeauthor{Garrington1999} had a detection sensitivity
between 1\,mJy and 2\,mJy and targeted FIRST sources with 1.4\,GHz peak flux
densities above 10\,mJy, whereas \citeauthor{Deller2014} targeted all kinds of
FIRST sources without any preselection at a varying detection sensitivity. In
order to compare their detection fraction with that of
\citeauthor{Garrington1999}, \citeauthor{Deller2014} cut their catalogue to
FIRST sources with $S_{1.4\,\mathrm{GHz}} > 10$\,mJy\,beam$^{-1}$ and to the
VLBI detection sensitivity from \citeauthor{Garrington1999}.
\citeauthor{Deller2014} found a detection fraction of 36\% in that subsample, in
agreement with the number from \citeauthor{Garrington1999}.\par

We followed the approach from \citet{Deller2014} and compiled a subsample of our IFRS
sample by including only those sources with an arcsec-scale 1.4\,GHz flux density above
10\,mJy\,beam$^{-1}$, ending up with 56~IFRS. Setting our detection sensitivity
to 1.5\,mJy, we would have detected 25~($45^{+7}_{-6}$\%) out of these 56~IFRS,
i.e.\ a slightly higher fraction than those from \citet{Garrington1999} and
\citeauthor{Deller2014}. Using a \textit{Fisher's exact test}~(e.g.\
\citealp{Wall2012}), we found a probability of 0.88 that our sample has a
higher VLBI detection fraction than the sample from \citeauthor{Deller2014}.\par

It is known that the radio source population at 1.4\,GHz with flux densities
above 1\,mJy consists almost exclusively of AGNs~(e.g.\ \citealp{Condon2012},
Fig.~11). This implies that the VLBI-observed subsamples from
\citet{Garrington1999} and \citet{Deller2014} and the sample presented in this
work---all cut to 10\,mJy and matched to the same sensitivity as discussed
above---contain virtually only AGNs. Thus, based on the numbers given above, we
find a tendency of a higher VLBI detection fraction for IFRS compared to the general
AGN population.\par

\citet{Collier2014} discarded all objects from their IFRS catalogue which provided a spurious WISE counterpart
to one of the radio lobes. It is unclear whether this selection criterion can
explain the higher VLBI detection fraction of IFRS compared to the general AGN
population. Apart from that, the only difference in selecting the objects of the
general AGN sample and the IFRS sample is the application of the IFRS
selection criteria from \citet{Zinn2011} mentioned in Sect.~\ref{IFRSdefinition}.\par

\begin{figure}
	\centering
		\includegraphics[width=\hsize]{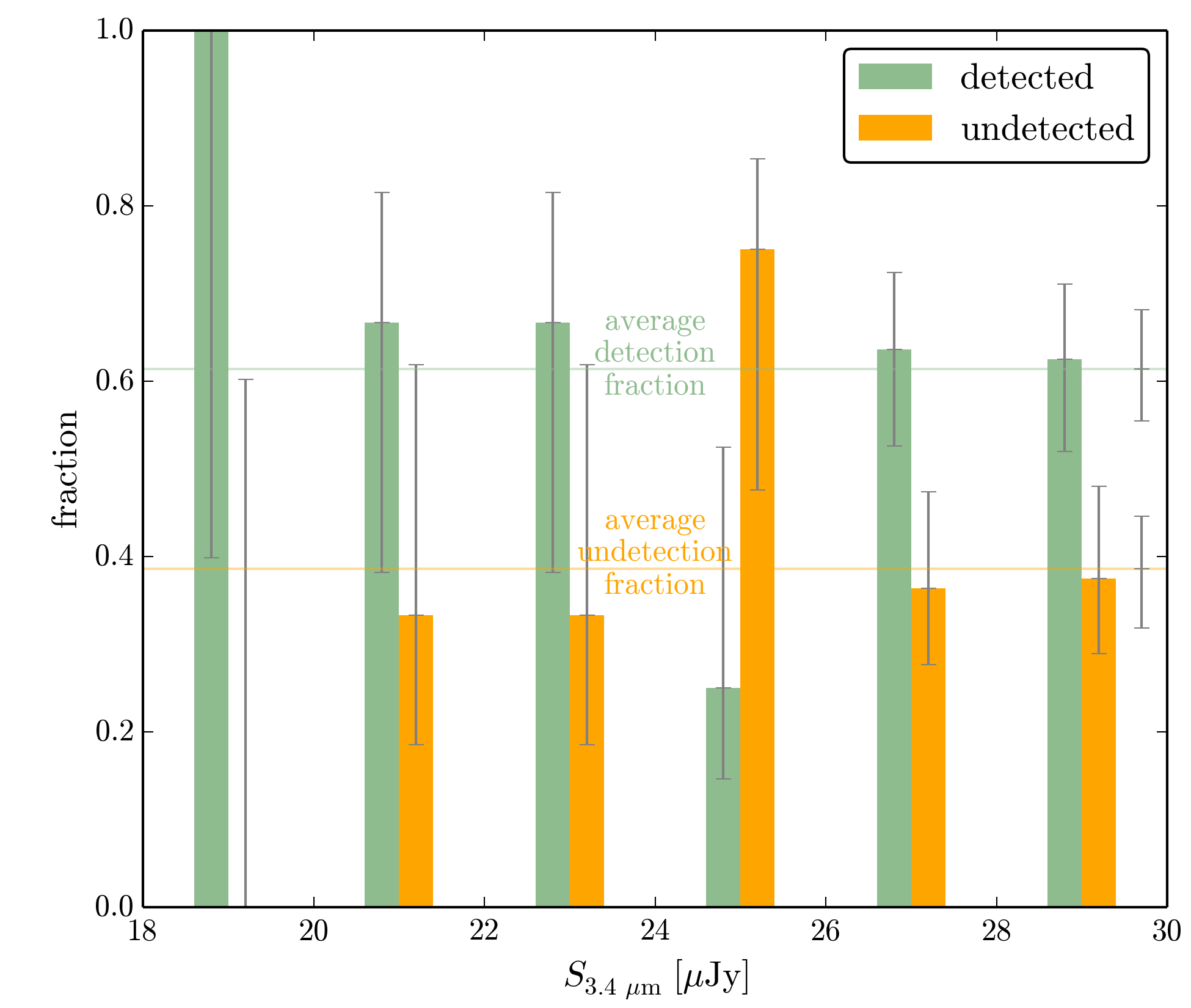}
		\caption{Histogram of the VLBA detection and non-detection fraction,
		binned by the flux density at 3.4\,$\mu$m. The
		horizontal lines show the overall detection and non-detection fraction
		of the observations presented in this work. $1\sigma$
		confidence intervals are shown in grey.}
	\label{fig:histogram_3p4umbins}
\end{figure}

Figure~\ref{fig:histogram_3p4umbins} shows the VLBA detection fraction binned in
the 3.4\,$\mu$m flux density for our sample of IFRS.
We do not find any evidence for a dependence of the VLBI detection fraction on
the 3.4\,$\mu$m flux density. We also tested the detection fraction against the
arcsec-scale 1.4\,GHz flux density, the radio-to-IR flux density ratio, and the
WISE colour [W1-W2] and found no significant correlation.\par

Compact radio cores are detected in eight~($73^{+9}_{-16}$\%) of the eleven IFRS
with SDSS counterpart and in 24~($57^{+7}_{-8}$\%) of the 42~IFRS covered
by SDSS without SDSS counterpart. However, we do not consider this a significant
difference (a)~because these two subsamples are not flux-complete, (b)~because of the
non-uniform sensitivity of SDSS, and (c)~because SDSS is biased towards
detecting blue objects as discussed above.\par

Three out of the eleven IFRS with SDSS counterpart are classified as ``galaxy''
in SDSS. We detected two~($67^{+14}_{-28}$\%) of these three galaxy-type IFRS in
our VLBA observations. The other eight IFRS with SDSS counterpart are classified
as ``star'' in SDSS. We note that this photometry-based classification is based
on the extension of the object, i.e.\ objects classified as star are
point-like, whereas extended objects are classified as galaxy. Out of these
eight IFRS classified as star, we detected six ($75^{+9}_{-19}$\%) in our VLBA
observations. \citet{Deller2014} found a higher VLBI detection fraction for
sources classified as star-like in SDSS. Our results are in agreement with this finding.\par

\subsubsection{Dependence of the VLBI detection fraction on radio properties}
\label{detection_fraction_radiospectrum}

Gigahertz peaked spectrum~(GPS) and compact steep spectrum~(CSS) sources are
very compact AGNs and expected to be the earliest phases in the evolution of
AGNs. GPS sources have a turnover frequency of around 1\,GHz and are usually
less than 1\,kpc in size, whereas CSS sources are more extended with a size of a
few kpc or a few tens of kpc. CSS sources are named for their steep radio
spectra ($\alpha \leq -0.5$; e.g.\ \citealp{Randall2011}).\par

Based on data at 6\,cm, 20\,cm, and 92\,cm, \citet{Collier2014} classified 124
of their IFRS as CSS sources and 32 as GPS sources. Out of the 57~IFRS observed
with the VLBA, five IFRS were classified as CSS sources and two as GPS sources.
In our VLBA observations, we detected four out of five IFRS which were
classified as CSS sources and both IFRS which were classified as GPS sources so
that CSS/GPS sources have a higher detection rate ($85^{+5}_{-21}$\%) than the
non-classified sources ($58^{+6}_{-7}$\%).\par

\citet{Collier2014} used the lower-resolution data of NVSS for the
flux densities in their IFRS catalogue. They also listed the number of sources
in the higher-resolution FIRST survey associated with the NVSS source. IFRS
detected with more than one FIRST component are clearly extended radio
galaxies and not GPS or CSS sources. Out of the 57 IFRS observed with the
VLBA, 47 are associated with exactly one FIRST source, while ten IFRS are
associated with two or three FIRST sources. Out of these ten IFRS with two or three FIRST counterparts, we detected
one ($10^{+17}_{-3}$\%) in our VLBA observations, whereas 34 ($72^{+5}_{-7}$\%)
out of 47 IFRS with exactly one FIRST counterpart were detected with the VLBA.
We found a statistically significantly higher VLBA detection fraction for
IFRS with exactly one FIRST counterpart compared to the detection fraction
of IFRS with more than one FIRST counterpart.\par

We also compared our VLBA detection fraction of IFRS with exactly one
FIRST counterpart to the detection fraction found by \citet{Deller2014} for the
general radio source population. As described above, we matched the arcsec-scale
radio flux density and the VLBA detection sensitivity to 10\,mJy\,beam$^{-1}$
and 1.5\,mJy\,beam$^{-1}$, respectively. We found a detection fraction of
$54^{+7}_{-7}$\% for those IFRS with exactly one FIRST counterpart, compared to
a detection fraction of 36\% for the general radio source population above
10\,mJy\,beam$^{-1}$ measured by \citeauthor{Deller2014}. Thus, our sample of
IFRS with exactly one FIRST counterpart is statistically different to the
general radio source population, based on a probability of 0.01 in a
\textit{Fisher's exact test}~(e.g.\ \citealp{Wall2012}) that the two
samples are taken from the same parent population.\par

\subsection{Compactness}
\label{compactness}
We determined ratios of the integrated mas-scale flux density in the VLBA
observations at 1.4\,GHz to the integrated arcsec-scale flux density in NVSS at
1.4\,GHz to fall between 0.86 and 0.014 for the IFRS detected in our VLBA
observations. On average, this ratio, which we refer to as compactness, was
$0.33\pm 0.23$.
\begin{figure}
	\centering
		\includegraphics[width=\hsize]{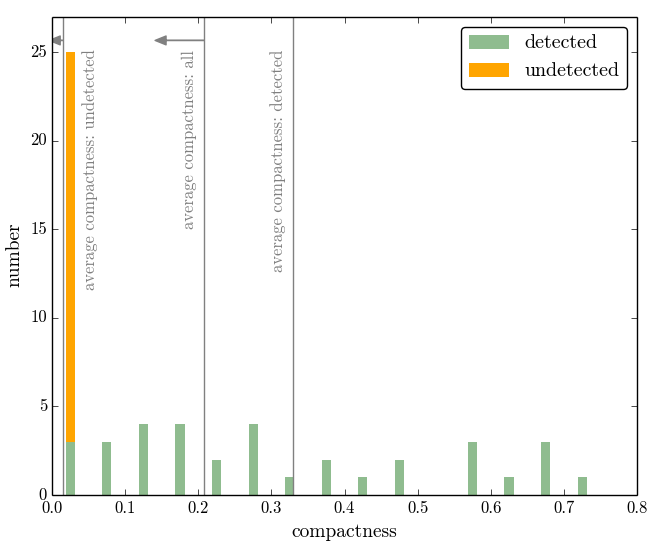}
		\caption{Histogram of the VLBA detections and non-detections, binned in
		the compactness at 1.4\,GHz. Compactness is defined as the ratio of
		mas-scale flux density to arcsec-scale flux density. The vertical lines
		show the mean compactness---from left to right---of all VLBA-undetected IFRS,
		of all VLBA-observed IFRS, and of all VLBA-detected IFRS, respectively. The
		former two lines represent upper limits as indicated by the horizontal
		arrows.}
	\label{fig:histogram_compactnessbins}
\end{figure}
Figure~\ref{fig:histogram_compactnessbins} shows the number of detections binned
by the compactness.\par

The mean compactness of our detected IFRS of $0.33\pm 0.23$ is lower than that
of the two former VLBI detections of IFRS where \citet{Norris2007} and
\citet{Middelberg2008IFRS_VLBI} detected 88\% and 58\%, respectively, of the
arcsec-scale flux density. This discrepancy may be due to small-number
statistics or because our fluxes are measured on smaller scales than those from
\citeauthor{Norris2007} and \citeauthor{Middelberg2008IFRS_VLBI}. The restoring beam of the LBA observations
presented by \citeauthor{Middelberg2008IFRS_VLBI} was
$51.7\,\mathrm{mas}\,\times 23.6\,\mathrm{mas}$, i.e.\ 17~times larger compared
to the median beam of $14.3\,\mathrm{mas}\,\times 4.7\,\mathrm{mas}$ in our VLBA
observations. \citeauthor{Norris2007} did not image their $uv$~data because of
the poor $uv$~coverage, but their angular resolution was similar to that of
\citeauthor{Middelberg2008IFRS_VLBI}. Therefore, a lower fraction of detected
flux in our VLBA observations could be expected.\par

\subsubsection{Dependence of the compactness on the redshift}
\label{compactness_redshift}

In the following, we tested our data against a potential correlation between
redshift and compactness. SDSS DR10 provides spectroscopic redshifts for
three out of our 57~IFRS. Two ($z=2.11$ and $z=2.62$) of those are
undetected and one ($z=2.55$) is detected in the VLBA observations. However,
IFRS F0912 at $z=2.62$ was observed for only $\sim 25$\,min with the VLBA,
resulting in a sensitivity only half that of the other sources. Since the number
of objects in this subsample is too low to test our data, we extended our
subsample by including those IFRS with photometric redshifts presented in
Sect.~\ref{opticalproperties_redshifts}.\par

\begin{figure}
	\centering
		\includegraphics[width=\hsize]{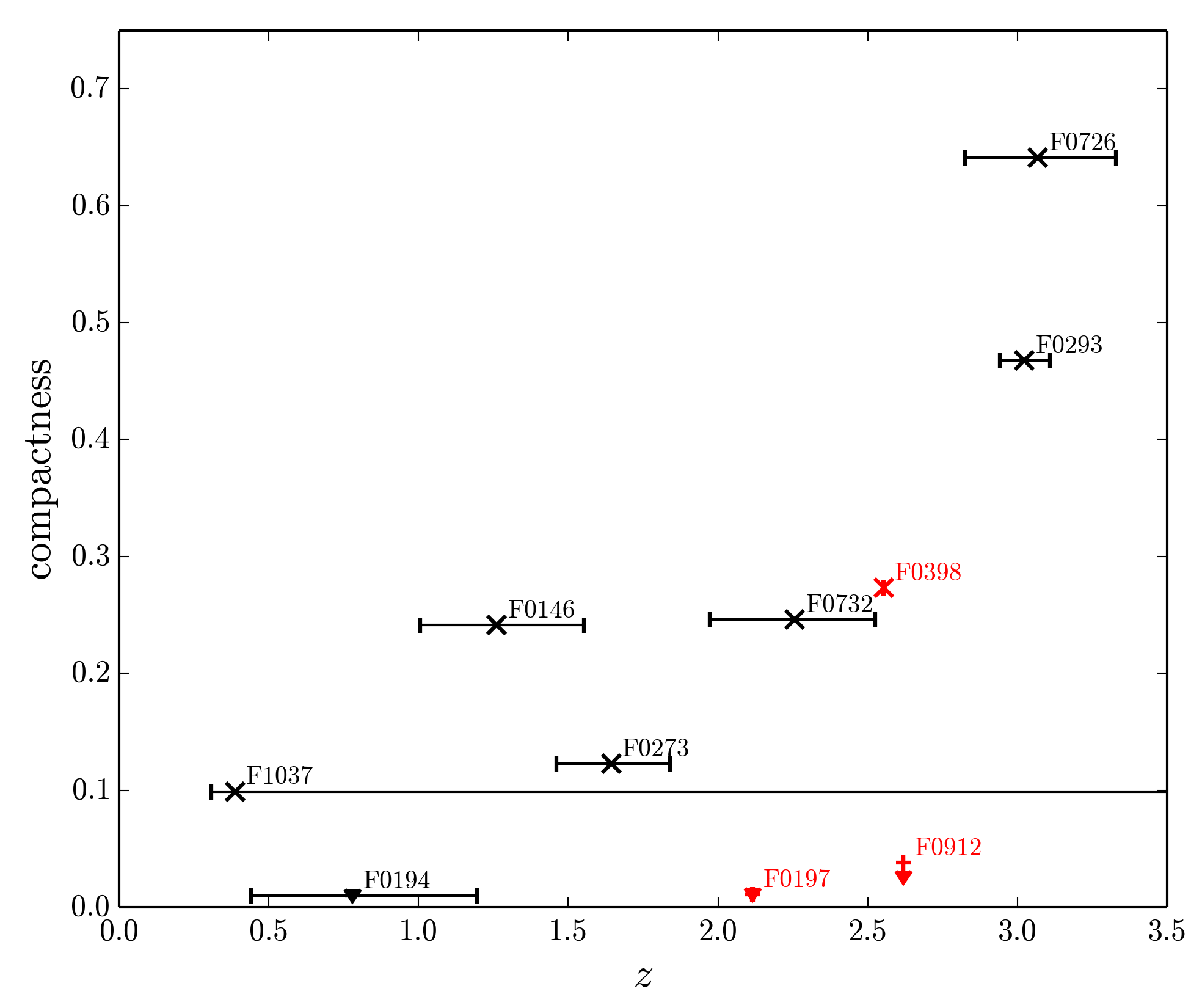}
		\caption{Compactness as a function of redshift for those IFRS with
		redshift information.
		VLBA-detected IFRS are shown by crosses and the $6.75\sigma$ upper limits on
		the compactness of VLBA-undetected IFRS are shown by arrows. Three
		spectroscopic (red markers) and seven photometric (black markers) redshifts were used in this
		analysis. Note that the errors on the compactness are around 20\%.}
	\label{fig:compactness_vs_z}
\end{figure}

Figure~\ref{fig:compactness_vs_z} shows the compactness as a function of the
redshift for all ten VLBA-observed IFRS with redshift information.
The data do not provide compact objects at low redshifts, whereas compact
objects were found at higher redshifts. The data suggest a possible
correlation between compactness and redshift for IFRS. We tested the data using
a \textit{Spearman rank correlation test}~(e.g.\ \citealp{Wall2012})
and found a correlation coefficient between 0.66 and 0.52, indicating a positive correlation between redshift and
compactness. A correlation coefficient of $+1$ and $-1$ represents an ideal
correlation and anticorrelation, respectively, whereas an uncorrelated data set
is represented by a coefficient of 0. In our case, the probability that the two
parameters are uncorrelated is between 0.019 and 0.063. The margin arises from
the unknown compactnesses of the VLBA-undetected sources for which only upper
limits are known. We determined this margin using a permutation test. Based on
the strong positive correlation coefficients, we suggest a correlation between
compactness and redshift for our sample of IFRS. When considering only the
VLBA-detected sources, we found a correlation coefficient of 0.96 and a
probability of $4.5\times 10^{-4}$ that the parameters are uncorrelated.
We cautiously note that the putative positive correlation seems to be
mainly based on the two highest-redshift IFRS in Fig.~\ref{fig:compactness_vs_z}
which might be outliers. Therefore, we emphasise that this suggested correlation
needs further testing.\par

\subsubsection{Dependence of the compactness on the 1.4\,GHz flux density}
\label{compactness_radioflux}

\begin{figure}
	\centering
		\includegraphics[width=\hsize]{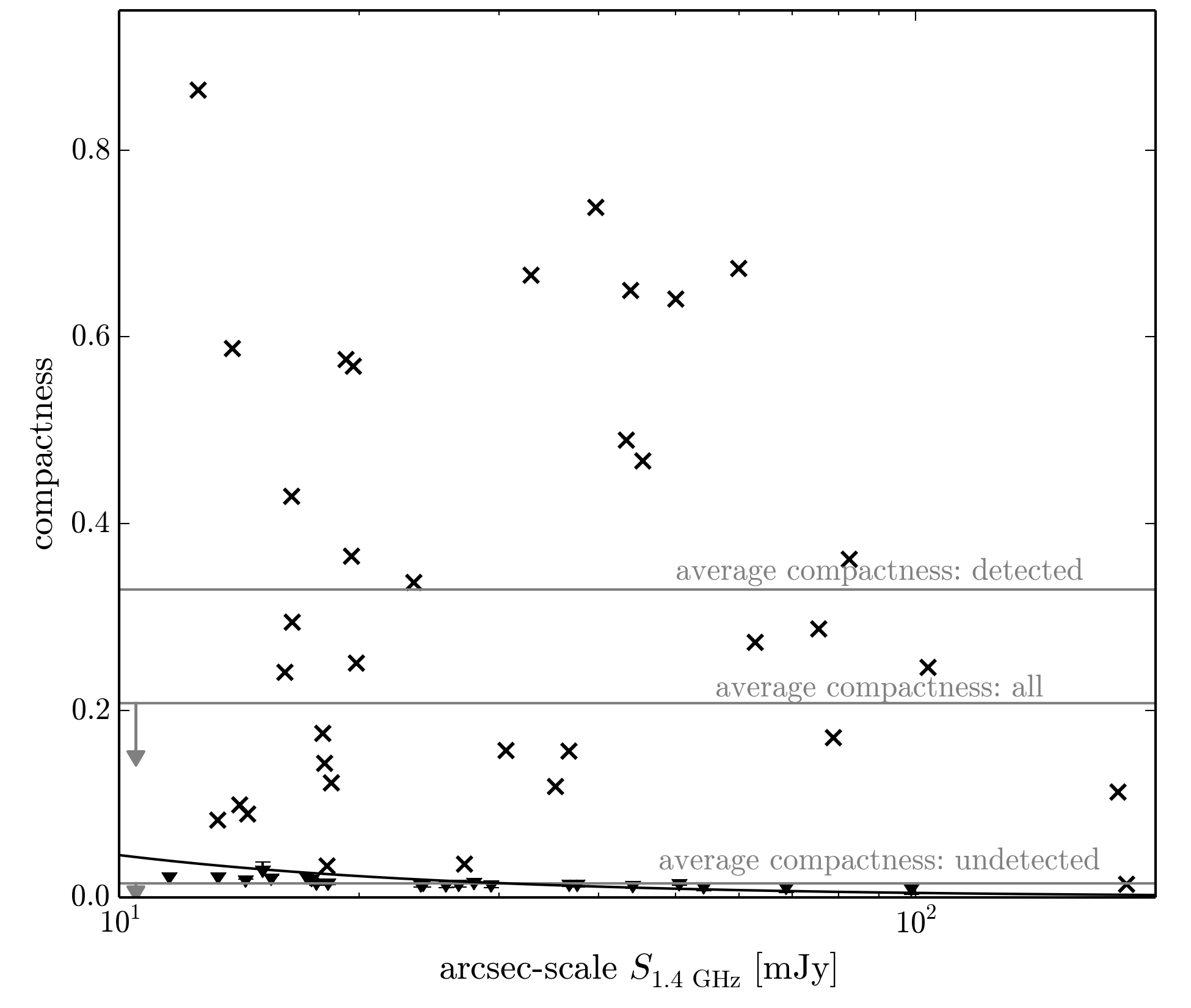}
		\caption{Compactness as a function of the arcsec-scale 1.4\,GHz flux density.
		VLBA-detected IFRS are shown by crosses and the $6.75\sigma$ upper limits on
		the compactness of VLBA-undetected IFRS are shown by black arrows. The grey
		horizontal lines represent---from top to bottom---the mean compactness of all VLBA-detected IFRS, of all
		VLBA-observed IFRS, and all VLBA-undetected IFRS, respectively. The latter
		ones represent upper limits as indicated by grey arrows. The black line shows
		the minimal detectable compactness depending on the arcsec-scale flux density, based on an mean
		detection sensitivity of 450\,$\mu$Jy in our VLBA observations.}
	\label{fig:compactness_vs_S14GHz}
\end{figure}
Figure~\ref{fig:compactness_vs_S14GHz} shows the compactness as a function of
the arcsec-scale 1.4\,GHz flux density and includes detections and upper limits
for the non-detections. We did not find compact radio-bright IFRS, whereas
compact radio-faint IFRS are common in our sample. If we divide our sample at an
arcsec-scale flux density of 60\,mJy, we find twelve ($25^{+7}_{-5}$\%) sources
with compactnesses above 0.4 and 36 ($75^{+5}_{-7}$) sources with compactnesses
below 0.4 in the fainter subsample. At arcsec-scale flux densities above
60\,mJy, we find nine ($100_{-17}$\%) sources with compactnesses below 0.4 and
no ($0^{+17}$\%) source with a compactness above 0.4. This is in agreement with
results from \citet{Deller2014}, who found a statistically significant
anti-correlation between compactness and arcsec-scale 1.4\,GHz flux density in
their sample of randomly selected radio sources. We used a \textit{Spearman rank
correlation test}~(e.g.\ \citealp{Wall2012}) to test for putative
correlations.
However, because of the significant fraction of upper limits in this plot, we can only narrow down the
correlation coefficient to a rather broad range. We found that the correlation
coefficient is between 0.39 and $-0.15$. Based on this test, we cannot exclude
either a positive or a negative correlation or a decorrelation.\par

\subsection{Individual sources}
\label{individual_sources}
In the following, we discuss three individual sources which are of particular
interest.

\subsubsection{F0398}
\label{F0398}
The only VLBA-detected IFRS with spectroscopic redshift is F0398 at $z=2.55$,
showing an arcsec-scale 1.4\,GHz integrated flux density of 62.9\,mJy.
This corresponds to a $K$-corrected 1.4\,GHz rest-frame luminosity of $2.3
\times 10^{27}$\,W\,Hz$^{-1}$, using the radio spectral index $\alpha = -0.72 $
between 20\,cm and 92\,cm from \citet{Collier2014}. In our VLBA observations,
the source---shown in Fig.~\ref{fig:F0398}---is slightly resolved with a peak
flux density of 13.7\,mJy\,beam$^{-1}$ and an integrated flux density of 17.2\,mJy,
corresponding to a luminosity of $6.3 \times 10^{26}$\,W\,Hz$^{-1}$ on scales
smaller than $146\,\mathrm{pc}\times 43\,\mathrm{pc}$. Based on this
luminosity, F0398 can be classified as
Fanaroff-Riley~(FR; \citealp{FanaroffRiley1974}) type~II. The source has a
compactness of 0.283.\par

\subsubsection{F0030}
\label{F0030}
\begin{figure}
	\centering
		\includegraphics[width=\hsize]{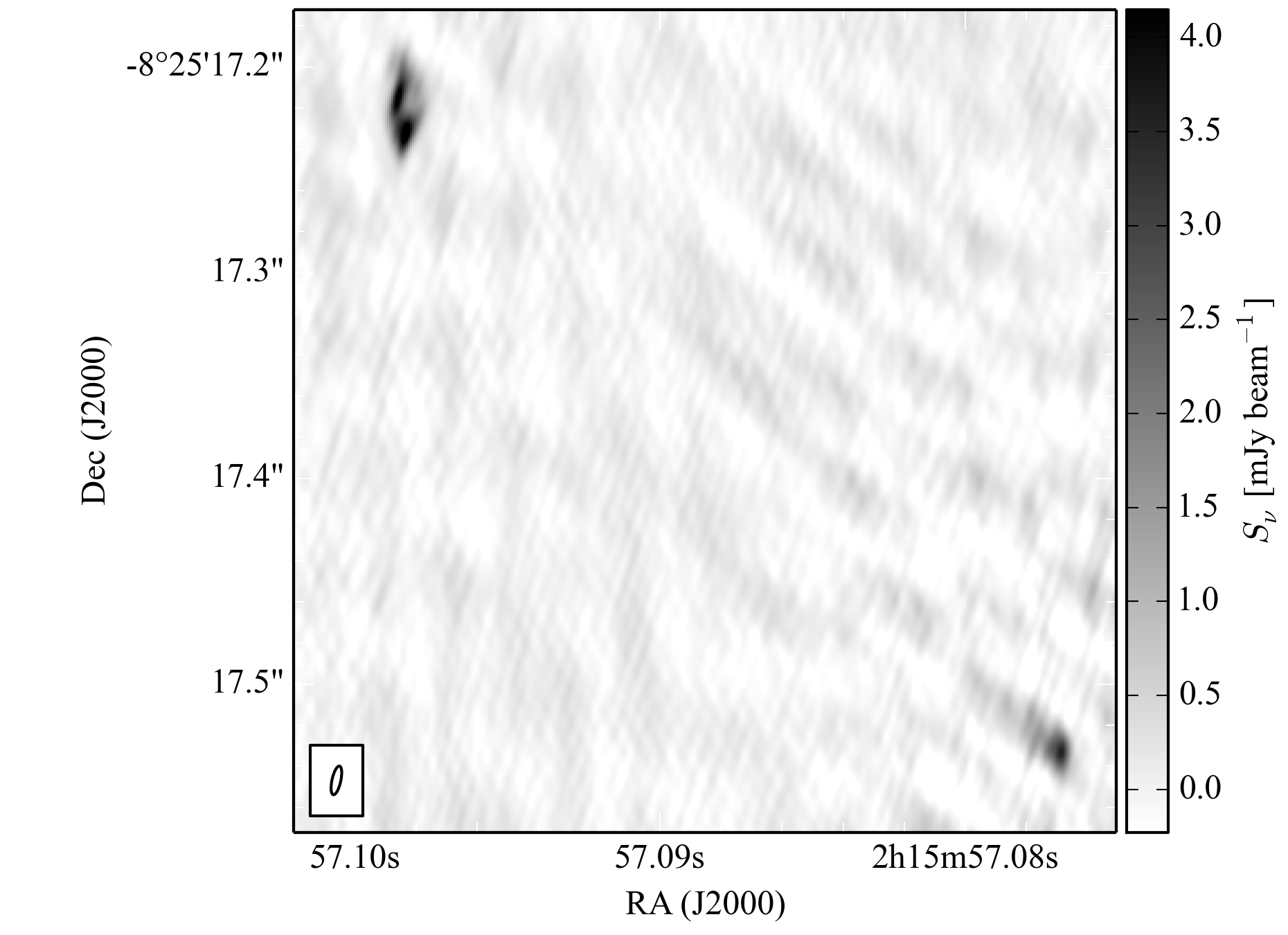}
		\caption{VLBA map of IFRS~F0030. The brighter component (upper left)
		has a complex morphology and a flux density of 21.8\,mJy.
		The second component (lower right) has a flux density of 6.0\,mJy and is
		separated by $442.1^{+0.3}_{-0.3}$\,mas, corresponding to a distance between
		1.7\,kpc and 3.8\,kpc at any reasonable redshift.}
	\label{fig:F0030}
\end{figure}
A particularly interesting source is F0030 which has two spatially separated
components in the VLBA map shown in Fig.~\ref{fig:F0030}. The first, brighter
component shows a mas-scale flux density of 21.8\,mJy ($\mathrm{S/N} = 55$) and
is spatially resolved with a complex morphology which is unique in our
observations. The second component shows a flux density of 6.0\,mJy
($\mathrm{S/N} = 46$) and is separated by $442.1^{+0.3}_{-0.3}$\,mas. The
linear distance between both components is between 1.7\,kpc and 3.8\,kpc at any
redshift in the range $0.5\leq z \leq 12$.\par

In order to obtain a rough spectral index of the components, we separately
imaged the four lower-frequency basebands and the four higher-frequency
basebands centred at 1.380\,GHz and 1.508\,GHz, respectively. Fluxes were
measured as described in Sect.~\ref{calibration_imaging}. We obtained spectral
indices of $-1.2\pm 1.2$ and $-1.2\pm 0.6$ for component~1 and 2,
respectively.\par

\subsubsection{F0257}
\label{F0257}
\begin{figure}
	\centering
		\includegraphics[width=\hsize]{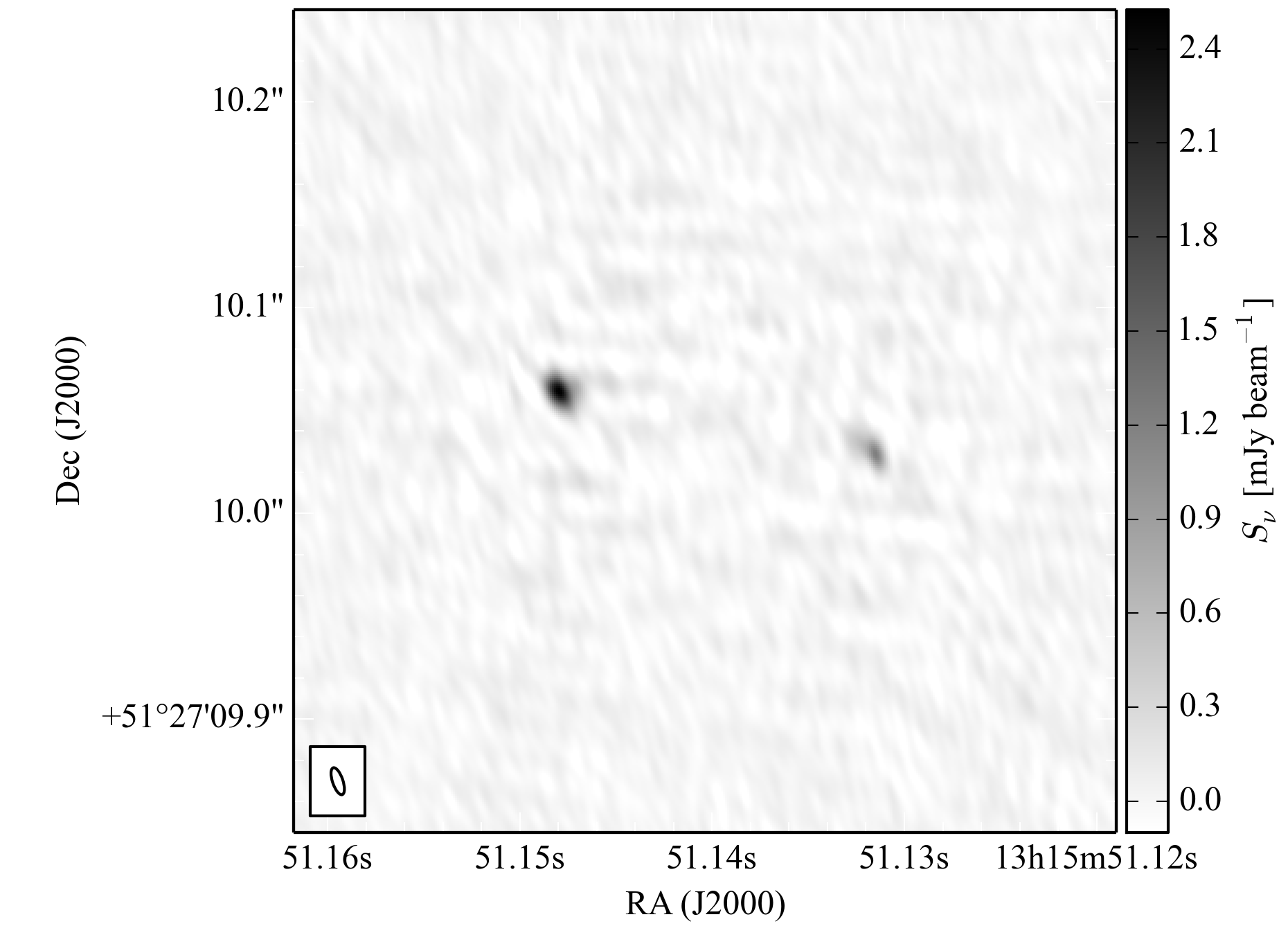}
		\caption{VLBA map of IFRS~F0257. The brighter component (left)
		has a flux density of 5.7\,mJy. The second
		component (right) has a flux density of 2.1\,mJy and is
		separated by $155.5^{+0.2}_{-0.2}$\,mas, corresponding to a distance between
		0.6\,kpc and 1.3\,kpc at any reasonable redshift.}
	\label{fig:F0257}
\end{figure}
Source F0257---shown in Fig.~\ref{fig:F0257}---consists of two individual
components. While the brighter component shows a flux density of 5.7\,mJy
($\mathrm{S/N} = 77$), the weaker component is $155.5^{+0.2}_{-0.2}$\,mas
distant and shows less than half of the other component's flux density
(2.1\,mJy\,beam$^{-1}$, $\mathrm{S/N} = 36$). This angular distance corresponds to a linear distance between 0.6\,kpc and 1.3\,kpc.
We measured mas-scale radio spectral indices between 1.380\,GHz and 1.508\,GHz
of $-0.3\pm 0.4$ and $0.2\pm 0.9$, respectively, for component~1 and 2.
\citet{Collier2014} measured an arcsec-scale radio spectral index of $-0.6$
between 6\,cm, 20\,cm, and 92\,cm.\par

\subsubsection{Could double component sources be a coincidence?}
\label{doublesources}

Sources F0030 and F0257 have both two compact components separated by 445\,mas
and 155\,mas, respectively. In the following, we estimate the probability that
two close by components are unrelated background sources. NVSS found 1.8~million
radio sources in an area of $3\times 10^4\,\mathrm{deg}^2$, out of which less
than 50\% have compact cores detectable in VLBI observations~\citep{Deller2014}.
This corresponds to a sky density of $1.8\times 10^{-6}\,\mathrm{arcsec}^{-2}$.
The probability of finding an additional unrelated source at a given position in
an area of $0.5\,\arcsec \times 0.5\,\arcsec$ is therefore of the order of
$5\times 10^{-7}$. Thus, we can effectively rule out any chance that the two
components found both in F0030 and F0257 are physically unrelated.\par

\section{Discussion}
\label{discussion}

Our observations increase the number of VLBI-detected IFRS from 2 to 37. Based
on our detection fraction of $61^{+6}_{-7}$\% and a reasonably large sample
size, we find strong evidence that most---if not all---IFRS contain AGNs. This
result confirms earlier studies by e.g.\ \citet{GarnAlexander2008},
\citet{Middelberg2011}, and \citet{Herzog2014}, who suggested compact cores in
IFRS based on SED modelling, radio-to-IR flux density ratios, and emission lines
in optical spectra. With higher sensitivity, we would have most likely detected
more sources in our VLBA observations.\par

We also tested our data for different potential correlations in
Sect.~\ref{analysis}. Although not all of them are statistically significant, in
the following, we explore the astrophysical consequences if these results are
confirmed by subsequent observations and describe how these hypotheses can be
tested.\par

In Sect.~\ref{compactness_radioflux}, we found a tendency that radio-brighter
IFRS are less compact. \citet{Deller2014} found the same behaviour when testing
the general AGN population and argued that this anti-correlation might be
explained by Doppler boosting effects as presented by \citet{Mullin2008}.
\citeauthor{Mullin2008} studied a complete sample of narrow-line and broad-line
radio galaxies and found an anti-correlation between radio luminosity and core
prominence. They argued that higher-luminosity sources have higher boosting
factors, associated with narrower boosting solid angles and a higher fraction of
sources for which a Doppler suppressed core is seen. In contrast,
lower-luminosity sources have lower boosting factors and wider solid angles,
corresponding to a lower core supression fraction and a higher compactness.\par

We argue that another factor might contribute to the observed behaviour in our
sample. It is known that AGNs evolve from very compact to extended objects by
forming and expanding jets, associated with an increasing total luminosity.
Namely, GPS sources are most compact and evolve into CSS sources and finally
into the largest radio galaxies, FR type I or II~(e.g.\
\citealp{ODea1998,Snellen1999,Fanti2009a}). Following this sequence, we would
expect younger AGNs to be more compact than old AGNs.\par

We note that both effects might overlap since they both predict a lower
compactness at higher luminosities. Based on the slightly higher VLBI detection
fraction of IFRS---which is expected to be a result of higher
compactnesses---compared to the general AGN population as found in
Sections~\ref{detection_fraction_AGN} and
\ref{detection_fraction_radiospectrum}, we suggest that IFRS are on average
younger than the general AGN population. This would be in agreement with results
by \citet{Middelberg2011} and \citet{Collier2014} who found some IFRS to be GPS
and CSS sources. The higher VLBI detection fraction for IFRS classified as GPS
and CSS sources compared to non-classified IFRS reported in
Sect.~\ref{detection_fraction_radiospectrum} also agrees with this
reasoning.\par

We suggest that---at least some---IFRS are young AGNs. However, the
sample presented by \citet{Collier2014} contains IFRS with different
characteristics. IFRS, which are associated with more than one FIRST source, are
clearly no GPS or CSS sources but extended radio galaxies. Those sources would
be expected to be older and less compact than IFRS with exactly one FIRST
counterpart. We found evidence for this expected behaviour in
Sect.~\ref{detection_fraction_radiospectrum} based on a lower detection fraction
for IFRS that are extended in arcsec-resolution images. We therefore suggest
that these extended IFRS are on average older and more evolved than the
VLBA-detected IFRS.\par

We found lower mean compactnesses of our VLBA-detected IFRS compared to the
other two VLBI-detections of IFRS by \citet{Norris2007} and
\citet{Middelberg2008IFRS_VLBI}. In addition to the technical explanations given
in Sect.~\ref{compactness}, this discrepancy can also be explained by boosting
effects. The IFRS from \citeauthor{Norris2007} and
\citeauthor{Middelberg2008IFRS_VLBI} are radio-fainter than the IFRS analysed in
our work. Following the reasoning by \citet{Deller2014} that brighter objects
are more likely to be Doppler suppressed, lower compactnesses for our IFRS
compared to the fainter IFRS from \citeauthor{Norris2007} and
\citeauthor{Middelberg2008IFRS_VLBI} could be expected.\par

In Sect.~\ref{compactness_redshift}, we found a statistically significant
correlation between redshift and compactness, with higher-redshift IFRS being
more compact. Two arguments can explain this correlation. (a) Higher-redshift objects have a
tendency to be younger than low-redshift versions of the same class of object.
Combining the increasing luminosity of GPS and CSS sources with time and the
boosting-related argument of decreasing compactness with luminosity,
higher-redshift---and thus younger and fainter---IFRS would be expected to be
more compact. (b) At higher redshifts, IFRS are more likely to be located in gas-rich
environments as shown for high-redshift galaxies~\citep{Klamer2006}. The higher
gas density confines these objects and keeps them more compact.\par

These results are all in agreement with the scenario that IFRS are younger and
therefore less luminous compared to the general AGN population, resulting in
higher compactnesses and higher detection fractions. However, we stress that
this putative connection between the age of IFRS and their VLBI properties is
not statistically significant and needs further testing.\par

Alternatively, the slightly higher VLBI detection fraction of IFRS compared to
the general AGN population could also be explained by a higher dust content of
IFRS, making it harder for the jets to expand and resulting in a more compact
object. However, no evidence has been found that IFRS are obscured by dust.
On the contrary, \citet{Collier2014} and \citet{Herzog2014} argued that the
IR faintness of IFRS is not caused by dust extinction. The SEDs
resulting from our photometric redshift fitting presented in
Sect.~\ref{opticalproperties_redshifts} also indicate that at least some IFRS
are very blue and do not support the hypothesis that a significant fraction of IFRS
is associated with dusty galaxies.\par

Future observations will help to test the hypotheses made in this work.
In particular, additional VLBA observations---similar to the observations
presented in this work---scheduled for semester 15A will increase the sample
size of VLBI-observed IFRS, providing the basis for more robust tests.\par

We plan to match arcsec-resolution radio data at higher and lower
frequencies, enabling the measurement of spectral indices and turnover
frequencies of IFRS. This information will bring out a putative overlap between
IFRS and GPS/CSS sources and provide further insights into the evolutionary status of
IFRS (Herzog et al., in prep.).\par

Radio observations---exposing the intermediate-resolution morphologies
of IFRS---could discriminate between the two mechanisms which may be
responsible for the lower compactness of brighter objects: beaming and age. If beaming is the
predominant cause for this effect, the radio-fainter IFRS should mainly be
one-sided objects (core-jet) since a beaming effect is more likely to be seen
for fainter objects as discussed above. In contrast, radio-brighter IFRS would
be expected to show a more symmetric structure. However, if the anti-correlation
between compactness and arcsec-scale radio flux density is mainly driven by the
age of the objects, no difference in the morphology would be expected, although
radio-brighter objects should be larger.\par

Two IFRS stand out from our sample since they were found to be composed of two
mas-scale components, separated by a few hundred mas. There are four different
explanations for those sources which we now discuss.\par

(i) GPS/CSS double lobe sources: It is known that GPS and CSS double lobe
sources can appear as separate components in VLBI
observations~\citep{Snellen2003}. In that case, the two components would be
hot-spots in the two jets and steep spectral indices would be expected~(e.g.\
\citealp{Hovatta2014}), while the VLBI-undetected core of the source would be
between both components. GPS sources are usually smaller than
1\,kpc~\citep{ODea1998}, whereas CSS sources show extensions of a few to a few
tens of kpc~\citep{Fanti2009a,Randall2011}.\par

(ii) Compact core and jet of a GPS or CSS source: Related to the first scenario,
the two components could be the compact core of the AGN and a hot-spot in one
jet. In that case, the spectral indices of both components can be different.
While the hot-spot should provide a steep spectrum as discussed above, the core
component would most likely provide a flat spectrum. However, the core spectrum
could also be steep~\citep{Hovatta2014}.\par

(iii) Gravitational lensing: The appearence of two components can also be
explained by gravitational lensing~(e.g.\ \citealp{Porcas1998}). In that case,
the emission seen as two components would originate from one distant source whose emission is deflected
by the gravitation of a nearby object. Therefore, similar spectral indices for
the two components are expected. In F0030 and F0257, the two components are too
close to find a potential gravitational lensing effect in optical images.\par

(iv) Binary black hole: The two components could also be a binary black
hole~(e.g.\ \citealp{Burke2011}). In that case, the spectral indices of the
components could be flat or steep or mixed as discussed in (ii) for the compact core.\par

Based on the available data, we are not able to exclude any of those four
different explanations because of the large error bars on the spectral indices.
F0030 is unlikely to be a GPS source because of the linear size of more than
1.7\,kpc. Following the correlation between intrinsic peak frequency and linear
size of compact AGNs presented by \citet{ODea1997}, F0030 would be expected to
show a rest-frame turn-over frequency of less than a few hundred MHz.
Particularly, F0030 has different characteristics than the
high-redshift ($z=5.774$) steep spectrum source
J0836+0054~\citep{Petric2003,Frey2005}. This RL quasar shows arcsec-scale
properties ($S_\mathrm{1.4\,GHz} =$ 1.75\,mJy, $\alpha = -0.8$) similar to those
of IFRS. However, in contrast to F0030, J0836+0054 has a second arcsec-scale
radio component which was undetected in VLBI observations and is most
likely associated with a lower-redshift galaxy.\par

\section{Conclusion}
\label{conclusions}

We observed 57~IFRS with the VLBA and detected compact emission in 35 of them.
Based on these observations, we draw the following conclusions.
\begin{itemize}
  \item We tested the hypothesis that IFRS contain AGNs. Our observations
  finally confirm the suggested compact cores in the majority of---if not
  all---IFRS, establishing IFRS as a new class of AGN. Our data increase the
  number of VLBI-detected IFRS from 2 to 37.
  \item Our data suggest that radio-brighter IFRS are on average less
  compact. This finding agrees with the evolutionary scenario that young AGNs
  evolve by expanding jets, becoming radio brighter and less compact with time.
  However, boosting effects may play a role, too.
  \item We found a marginal tendency for IFRS to show a higher VLBI detection
  fraction compared to randomly selected sources with mJy arcsec-scale flux
  densities, i.e. mainly AGNs.
  In our sample, the detection fraction is higher for IFRS with exactly one
  FIRST counterpart and for IFRS classified as GPS and CSS sources.
  \item A statistically significant correlation between redshift and compactness
  was found in our data for IFRS with higher-redshift sources to be more compact. This is in
  agreement with higher-redshift sources being located in denser environments and having a
  tendency to be younger.
  \item Two sources show two components each, separated by between 0.4\,kpc and
  3\,kpc at any reasonable redshift. These components might be jet/jet or
  core/jet components of an AGN, a binary black hole, or arising from
  gravitational lensing.
\end{itemize}
All our findings are in agreement with the scenario that IFRS contain young AGNs
which are in an early stage of their evolution. Their jets are not yet formed or
expanded significantly, resulting in a very compact source. When evolving, the
jets expand and the total radio fluxes of the sources increase, while the
compactnesses decrease at the same time. We note that some IFRS already
formed jets as known from arcsec-resolution maps.\par

Our analyses in this work were limited because of the low number of objects in
relevant subsamples. Based on new data from VLBA observations of IFRS in
semester~15A and planned lower-resolution observations, we are
aiming at extending our study and further testing the hypotheses presented in this work.

\begin{acknowledgements}
ATD acknowledges support from an NWO Veni Fellowship.
The National Radio Astronomy Observatory is a facility of the National Science
Foundation operated under cooperative agreement by Associated Universities, Inc.
This work made use of the Swinburne University of Technology software
correlator, developed as part of the Australian Major National Research
Facilities Programme and operated under licence.

Funding for SDSS-III has been provided by the Alfred P. Sloan Foundation, the
Participating Institutions, the National Science Foundation, and the U.S.
Department of Energy Office of Science. The SDSS-III web site is http://www.sdss3.org/.

SDSS-III is managed by the Astrophysical Research Consortium for the
Participating Institutions of the SDSS-III Collaboration including the
University of Arizona, the Brazilian Participation Group, Brookhaven National
Laboratory, Carnegie Mellon University, University of Florida, the French
Participation Group, the German Participation Group, Harvard University, the
Instituto de Astrofisica de Canarias, the Michigan State/Notre Dame/JINA
Participation Group, Johns Hopkins University, Lawrence Berkeley National
Laboratory, Max Planck Institute for Astrophysics, Max Planck Institute for
Extraterrestrial Physics, New Mexico State University, New York University, Ohio
State University, Pennsylvania State University, University of Portsmouth,
Princeton University, the Spanish Participation Group, University of Tokyo,
University of Utah, Vanderbilt University, University of Virginia, University of
Washington, and Yale University.
\end{acknowledgements}


\bibliographystyle{aa} 
\bibliography{references} 

\end{document}